\documentclass[10pt, %
superscriptaddress,
 aps,
 prx,
floatfix,notitlepage
]{revtex4-2}

\usepackage{amsmath,amssymb, amsthm}
\usepackage[hidelinks=true,colorlinks=true,linkcolor=teal, citecolor=blue, bookmarks=true,breaklinks=true]{hyperref}
\usepackage[capitalize,nameinlink]{cleveref} 

\newtheorem{theorem}{Theorem}
\newtheorem{corollary}[theorem]{Corollary}
\newtheorem{lemma}[theorem]{Lemma}

\usepackage{tikz}

\usepackage{graphicx, overpic}
\usepackage{dcolumn}
\usepackage{bm, color,colortbl,xcolor}
\usepackage{float} 

\usepackage[T1]{fontenc}
\usepackage[utf8]{inputenc}

\usepackage{caption}
\usepackage[labelformat=simple]{subcaption}

\usepackage{mathtools}
\usepackage[normalem]{ulem}
\usepackage{todonotes}
\usepackage{soul}
\usepackage{macros_Li}
\usepackage[hidelinks=true,colorlinks=true,linkcolor=teal, citecolor=blue, bookmarks=true,breaklinks=true]{hyperref}
\usepackage[capitalize,nameinlink]{cleveref}
\setul{0.5ex}{0.3ex}
\setulcolor{red}
\usepackage{pifont}
\usepackage{cancel}
\usepackage{ulem}
\usepackage[inline]{enumitem}
\usepackage{braket}
\usepackage{physics}
\usepackage{comment}

\newcommand{\eps}{\varepsilon}
\def\polylog{\mathrm{polylog}}
\newcommand{\bigO}[1]{\mathcal{O}\left( #1 \right)}
\newcommand{\R}{\mathbb{R}}

\renewcommand{\d}{\mathrm{d}}

\newcommand{\Y}{\mathbf{Y}}

\newcommand{\Order}{\varsigma}
\newcommand{\IndexFunc}{\mathcal{I}}

\newcommand{\Z}{\mathbb{Z}}

\graphicspath{{./fig/}}
\begin{document}

\preprint{APS/123-QED}

\title{Quantum Algorithm for Subcellular Multiscale Reaction-Diffusion Systems}
\author{Margot Lockwood}
\affiliation{Pacific Northwest National Laboratory, Richland, WA 99352, USA}
\author{Nathan Wiebe}
\affiliation{Pacific Northwest National Laboratory, Richland, WA 99352, USA}
\affiliation{Department of Computer Science, University of Toronto, Toronto, Ontario M5S 2E4, Canada}
\affiliation{Canadian Institute for Advanced Research, Toronto, Ontario M5S 2E4, Canada}
\author{Connah Johnson}
\affiliation{Pacific Northwest National Laboratory, Richland, WA 99352, USA}
\author{Johannes Mülmenstädt}
\affiliation{Pacific Northwest National Laboratory, Richland, WA 99352, USA}
\author{Jaehun Chun}
\affiliation{Pacific Northwest National Laboratory, Richland, WA 99352, USA}
\author{Gregory Schenter}
\affiliation{Pacific Northwest National Laboratory, Richland, WA 99352, USA}
\author{Margaret S. Cheung}
\email{margaret.cheung@pnnl.gov}
\affiliation{Pacific Northwest National Laboratory, Richland, WA 99352, USA}
\affiliation{University of Washington, Seattle, WA 98195}
\author{Xiangyu Li}\email{xiangyu.li@pnnl.gov}
\affiliation{Pacific Northwest National Laboratory, Richland, WA 99352, USA}
\date{\today}

\begin{abstract} 
Computational modeling of cellular systems, where reactants are governed by biochemical equations and physical representations, requires extensive classical computing resources. These limitations significantly constrain the system size and spatiotemporal scales of simulations. A key challenge lies in the ``curse of dimensionality'', where the number of possible reaction terms grows exponentially with the number of species, and the computation of reaction rates involving many-body interactions becomes intractable in polynomial time on classical computers.
In this work, we introduce a quantum algorithmic framework designed to overcome these challenges, leveraging the architecture of quantum computing to simultaneously compute reaction rates and track the spatiotemporal dynamical evolutions of subcellular systems. We generalize the reaction-diffusion equation (RDE) for multiscale systems with arbitrary species count, encompassing higher-order interactions. Our approach achieves two principal quantum advantages: (i) an exponential quantum speedup in reaction-rate computation, contingent on the efficient preparation of polynomially accurate ground states on a quantum computer, and (ii) a quadratic scaling in spatial grid points and polynomial scaling in the number of species for solving nonlinear RDEs, contrasting sharply with classical methods that scale exponentially with the system’s degrees of freedom.
To our knowledge, this represents the first efficient quantum algorithm for solving multiscale reaction-diffusion systems. This framework opens the door to simulations of biologically relevant subcellular processes across previously inaccessible spatial and temporal scales, with profound implications for computational biology, soft matter physics, and biophysical modeling.

\end{abstract}

\maketitle

\section{Introduction}
Modeling biological processes within a cellular system is computationally intensive due to the complexity of simulating vast networks of macromolecular interactions 
\cite{Francis2025, Southern2008, papoian2014}. 
The dynamics of these interactions propagate throughout multiple scales both in time and in space  from microscopic biochemical reactions that drive subcellular processes to macroscopic phenomena such as signal transduction and metabolism of a cellular state
\cite{Gross2019, Sher2024}.  
Signal transduction is a type of emergent properties that enable cellular responses to environmental stimuli in concomitant with  metabolic adaptation that supports growth or nutrient transport \cite{Cheung2025}. These processes occur in highly crowded, polydisperse environments\cite{Cheung2019}, with reaction rates mediated by enzymes and requiring spatial organization at a subcellular level 
\cite{Johnson2021,Alberts2015}. 
Understanding these mechanisms through modeling of biochemical reactions with a physical representation in spatial content enables predictions of cellular states and emergent behavior
\cite{Chubukov2014,LUTHEYSCHULTEN2022}. 

It is promising to model the biological processes over multiple space and time scales with Reaction-Diffusion Equations (RDEs),  particularly for describing how cellular transport and interactions
\cite{Soh2010, LUTHEYSCHULTEN2013,papoian2014,Barsegov2022}. 
Accurately predicting the biological processes over a meaningful spatiotemporal span, however, requires a complete description of a cell state involving its shape, size, components, and intracellular reactions. The feat of the first four-dimensional (4-D) whole-cell kinetic simulation based on coupled RDEs was accelerated by powerful Graphics Processing Units (GPUs) for a minimal synthetic cell\cite{thornburg2022}.  It remains intractable to simulate a multicellular or complex organism with the same level of fine-grained descriptions due to the exponential growth of biochemical reaction terms among species as well as the challenges in combinatorial scaling with the degrees of freedom and spatiotemporal complexity. 
 
Particularly, an
accurate biochemical reaction rate governed by exact quantum
rate theory \citep{yamamoto_quantum_1960, miller_quantum_1983} involves Hamiltonians that are anharmonic and
nonadiabatic with high-dimensional degrees of freedom. It is notoriously nondeterministic-polynomial (NP) hard for classical 
approaches. 
Such a Hamiltonian 
  faces a fermionic sign problem that requires exponential
  simulation time
  \citep{ceperley_path_1995}, represented by a worst-case exponential barrier to classical
  sampling of many-body quantum
  dynamics \citep{troyer_computational_2005}.
  Quantum algorithms accelerate the calculation of appropriate rate problems by
  offering exponential speedups \citep{babbush_low-depth_2018, bauer_quantum_2020}.
  
  An equally challenging aspect of solving a nonlinear RDE system lies that it must be represented linearly to simulate on a quantum computer due to the linear nature of quantum mechanics. Whether a quantum advantage can be achieved depends on the efficiency of the linear mapping \citep{liu_efficient_2021} and fine prints of linearized RDE system \citep{aaronson_read_2015}.   
  We demonstrated efficient linear mapping and logarithmic scaling of system size in solving the highly nonlinear fluid dynamics on a quantum computer
 \citep{li_potential_2025}.
  Here, we applied the same Carleman linearization \citep{Carleman1932, steeb1990nonlinear, Kowalski1991, Baudouin2016Carleman} to the nonlinear RDE and then solve the linearized system.
 Compared to strongly nonlinear fluid dynamics  \citep{li_potential_2025}, the RDE system in this work was characterized by weak nonlinearly, and presented by a vast number of degrees of freedom. Most importantly, the latter are most favourable for quantum computing.        
 
\begin{figure}
   \centering
   \includegraphics[width=0.8\linewidth]{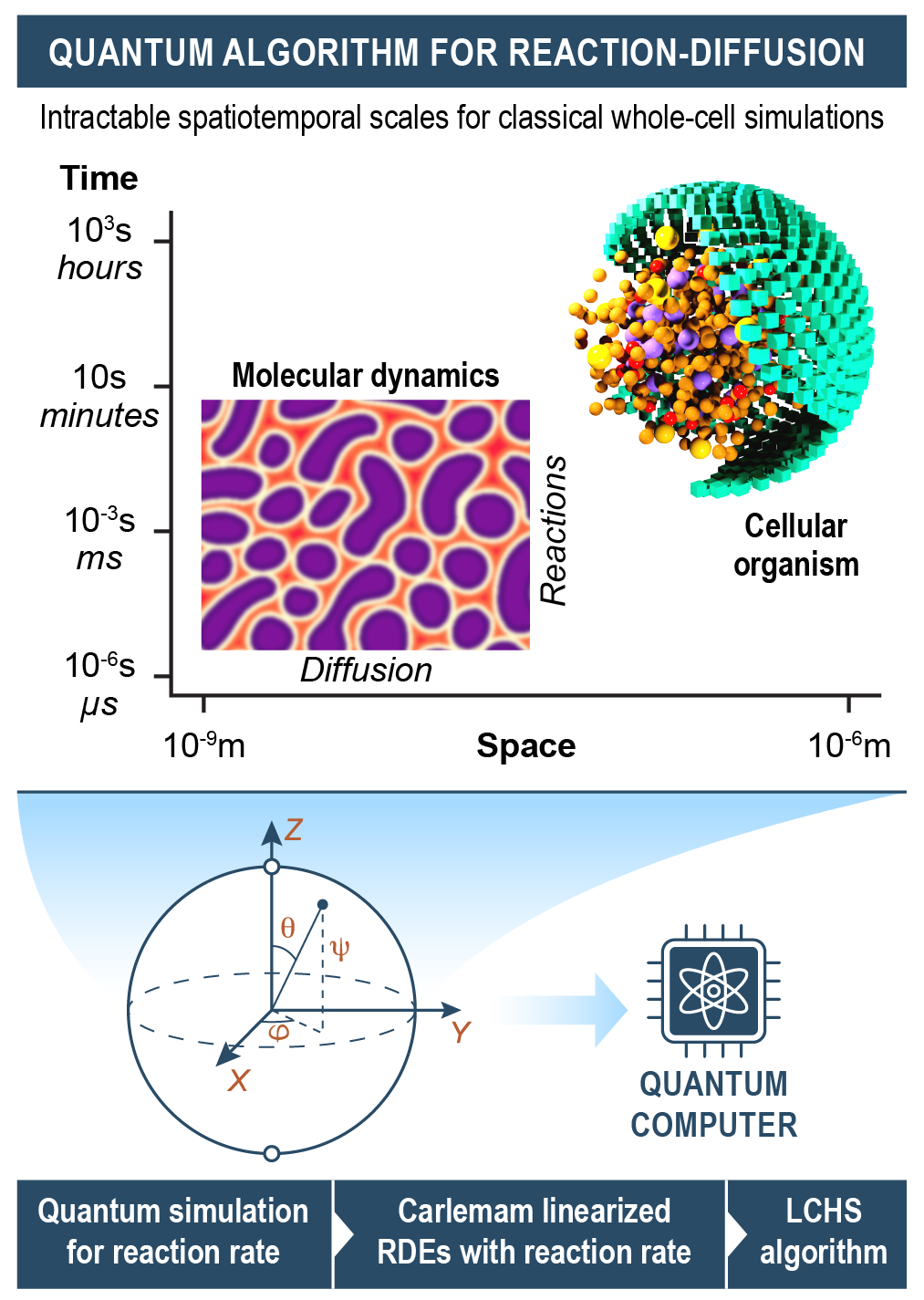}
   \caption{A schematic diagram for the development of an efficient quantum algorithm for reaction rate and the linear combination of Hamiltonian system (LCHS) for the coupled subcellular multiscale reaction-diffusion equations (RDE) relevant in whole-cell modeling in a broad spatiotemporal scale as described in \cref{tab:scales}.}
   \label{fig:overview}
 \end{figure}

In this work, we break the curse of dimensionality for both the biochemal reaction rates and the nonlinear RDE by exploring potential quantum advantage with explicit calculation of reaction rates (\cref{fig:overview}). For a use-case RDE example, we generalize the canonical Turing model of $S$-species and up to $\varsigma$-order reaction terms to harness the power of quantum computing in Hilbert space. The first breakthrough from this work is that we attained an exponential quantum speedup in calculating the reaction rates represented by the Erying equation.
In concomitant with calculating the reaction rate, we solved the RDEs on quantum computers. This unprecedented synergy saves us the exponential cost of preparing the initial states for the reaction rates from dynamic integration of simulating an RDE system. 

\begin{table}[t]
    \centering
    \renewcommand{\arraystretch}{1.3}
    \begin{tabular}{lccc}
        \toprule
        \textbf{Scale} & \textbf{Spatial Scale} & \textbf{Time Scale} & \textbf{Phenomena} \\
        Atomic & 0.1 -- 1 nm & Femtoseconds -- Nanoseconds &  Covalent bond breaking and forming \cite{Piris2024} \\
        Macromolecular & 1 -- 100 nm & Nanoseconds -- Microseconds & Molecular dynamics and interactions, protein folding \cite{Zwier2010} \\
        Cellular & 1 -- 100 $\mu$m & Microseconds -- Minutes & Gene regulation \cite{Meeussen2024}, protein diffusion \cite{Kumar2010}  \\
        Mesoscopic & 1 $\mu$m -- 1 cm & Seconds -- Hours & Molecular motor transport \cite{Ajaj2024}, Tissue formation \cite{Ho2024} \\
        Macroscopic & 1 cm -- 30 cm & Minutes -- Days & Physiological pharmacokinetics \cite{Seo2022} \\
    \end{tabular}
    \caption{Biological spatial and temporal scales for hierarchical multiscale biological modeling \cite{Barbulescu, fish2009multiscale}.}
    \label{tab:scales}
\end{table}

\section{Multiscale Dissipative Reaction-Diffusion Systems}

Whole-cell simulations typically utilize partial differential equations (PDEs) or chemical master equations to model the diffusion of chemicals through space and the reactions they participate in \cite{noireaux_development_2011}. To address cell modeling for quantum computers, we consider an extended Turing model with $S$ interacting concentrations. 
Turing’s seminal work on morphogenesis introduced a coupled system of reaction diffusion equations (RDEs) and showed how gradients of key signaling molecules-morphogens-can drive large-scale patterns of cell differentiation \cite{turing1952chemical}. Such models explain how local chemical interactions can generate macroscopic phenomena, including zebrafish stripe patterns \cite{Kondo2021} and cell-type patterning in biofilms \cite{Johnson2022}. RDEs capture the spatiotemporal evolution of cellular and chemical fields and are highly sensitive to initial and boundary conditions, which strongly influence pattern selection \cite{gierer1972theory}. 
Mathematically, the model is formed by a system of reaction-diffusion PDEs describing the dynamic changes in the morphogen concentration field with waves and diffusion-driven instabilities representing the changes in cell state. The biological mechanisms connecting the morphgen changes to cell differentiation were not initially suggested but the resulting framework has subsequently been used as the basis for more complex and predictive models \cite{Kondo2022}.

The original Turing model involves two morphogen chemical concentrations, an activator and an inhibitor, that are distributed over space and time. The dynamics of these concentrations are assumed to be governed by RDEs.  We present a generalized Turing model for \(S\) interacting concentrations, denoted as \(y_1(x,t), y_2(x,t), \dots, y_S(x,t)\), by:
\begin{align}\label{eq: Turing}
\frac{\partial y_i(x, t)}{\partial t} =\mathbf{R_i} \cdot \bigoplus_{j=0}^\varsigma \mathbf{Y}^{\otimes j} + D_i \nabla^2y_i(x, t), \quad \text{for } i=1, 2, \dots, S 
\end{align}
where $\mathbf{R_i=\begin{bmatrix} \mathbf{R_{0, i}} &\mathbf{R_{1, i}}&\cdots
\mathbf{R_{\varsigma, i}} \end{bmatrix}}$ are the vectors containing operators for the inhomogeneous, linear decaying, and nonlinear reaction terms at orders up to $\varsigma$ specified for each concentration. 
$\bigoplus$ denotes the direct sum of vectors or matrices, which in this case acts as a vertical concatenation. 
$\Y = [y_1, y_2, \ldots, y_S]^\top \in \R^S$ is the state vector.
$\Y^{\otimes j}$ denotes the $j$-th order Kronecker (tensor) power:
\begin{align}
\mathbf{Y}^{\otimes j} = [y_{i_1} y_{i_2} \cdots y_{i_j}]^\top \in \mathbb{R}^{S^j}
\end{align}
where $(i_1, i_2, \ldots, i_j) \in \{1,2,\ldots,S\}^j$ are ordered lexicographically. An example of $\Y^{\otimes 3}$ is given in \cref{eq:Y^3}. 
$D_i$ represents the diffusion coefficient for species $i$.
Define
$\mathbf{Y}^{\otimes0}:=[1]$ so $\mathbf{R_{0, i}} \cdot \mathbf{Y}^{\otimes 0}$
represents the inhomogeneous term for each equation. $\mathbf{R_{1, i}}$ collects
$S$-linear decaying terms. $\mathbf{R_{j, i}}$ contains $S^j$ higher order terms for $j\in 1, \dots, \varsigma$. 
The reaction terms vary based on the reaction dynamics of the physical system. The number of possible reaction combinations with $S$ concentrations grows exponentially due to the combinatorial nature of interactions among species. 

We use the polynomial-matrix form of \cref{eq: Turing} because it is more suitable for the application of quantum computing as will be discussed later, 
\begin{align}\label{eq:Turing-poly}
    \frac{d\mathbf{Y}}{d t}=\mathbf{F}_0 + \mathbf{F}_1\mathbf{Y} + \mathbf{F}_2 \mathbf{Y}^{\otimes 2}+\dots +\mathbf{F}_\varsigma \mathbf{Y}^{\otimes \varsigma}
\end{align}
where  $\mathbf{F}_j \in \mathbb{R}^{S \times S^j}, j \in \{0,\dots , \varsigma\}$ are \textit{time-independent} coefficient matrices.
The $0$-th order coefficient $\mathbf F_0 \in \R^S$ in \cref{eq:Turing-poly} represents
the inhomogeneous terms. The coefficient tensor for the linear terms of
\cref{eq:Turing-poly} represents the superposition of the linear decay rate $\mu_i$ and the diffusion
coefficient $D_i\, \text{for}\, i\in{1, 2, \dots, S}$, i.e.,
\begin{align}
\label{eq:F1}
\mathbf{F}_1 = 
\begin{bmatrix}
    D_1 \nabla^2  - \mu_1 & 0 & 0 & 0 & \cdots & 0 \\
    0 & D_2 \nabla^2  - \mu_2 & 0 & 0 & \cdots & 0\\
    \vdots & \vdots & \vdots & \vdots & \ddots \\
    0 & 0 & 0 & 0 & \cdots & D_S \nabla^2  - \mu_S\\
\end{bmatrix}. 
\end{align}

Reaction rates among $S$-species are represented by high-order terms ($\varsigma \ge 2$) in \cref{eq:Turing-poly}, the coefficient tensor ($\mathbf F_j$) of which is determined by the reaction rate constants $c_r:=c_{ij}$ where for $r\in \{1,\ldots R\}$ are the different reactions from products to reactants which we will at times denote $c_{ij}$ to refer to the reaction from a configuration $i$ to $j$.
Determining $c_{r}$ from first-principles is the key to accurately simulate \cref{eq:Turing-poly} in molecular dynamics \citep{mao_classical_2023}. This is classically challenging due to the exponential blowup of the dimension of reaction rate space, i.e., calculating $2^S$ reaction rates for a $S$-species reaction-diffusion system on just \textit{one} node.
The reaction rate  $c_{r}$ is determined by estimating the free energy between two states as discussed in depth in \cref{sec:reaction-rate}. 
To construct $\mathbf F_j$ ($j \ge 2$) for a general $S$-species mixed-order reaction system, we consider chemical reaction networks of the general form following the canonical mass action law \citep{feinberg_foundations_2019}:
\begin{align}\label{eq:general_reaction}
\sum_{i=1}^S \alpha_{r,i} y_i \xrightarrow{c_r} \sum_{i=1}^S \beta_{r,i} y_i \quad (r = 1, 2, \ldots, R)
\end{align}
where $\alpha_{r,i}, \beta_{r,i} \in \Z_{\geq 0}$ (non-negative integers) are stoichiometric coefficients (see examples in \cref{sec:poly-structure}) specifying the number of molecules of species $i$ as reactants and products for each reaction $r$, respectively. The stoichiometric coefficients encode atom, mass, and charge conservation constraints and define the stoichiometric matrix that relates reaction extents to species balances. $c_r > 0$ are rate constants.
The corresponding reaction rate is \citep{voit_150_2015}
\begin{align}\label{eq:rate_equations}
  \text{Reaction rate} & = \sum_{r=1}^R (\beta_{r,i} - \alpha_{r,i}) c_r \prod_{j=1}^S y_j^{\alpha_{r,j}} \quad (i = 1, 2, \ldots, S)\, , \\
  \label{eq:rd-stochi}
		       & = \sum_{\ell=1}^{S^\Order} F_\Order[i,\ell] \cdot [\mathbf{Y}^{\otimes \Order}]_\ell
\end{align}
which is the stoichiometric power representation of polynomial nonlinear terms in \cref{eq:Turing-poly}. Here, $[\mathbf{Y}^{\otimes \Order}]_\ell$ is the $\ell$-th component of the tensor product. 
The reaction rate is proportional to the product of reactant concentrations raised to their stoichiometric powers \citep{steinfeld_chemical_1999}.
For a reaction $r$, we define order $\Order_r = \sum_{i=1}^S \alpha_{r,i}$ (total number of reactant molecules that must simultaneously collide)
and molecular multiplicity vector $\boldsymbol{\alpha}_r = (\alpha_{r,1}, \alpha_{r,2}, \ldots, \alpha_{r,S}) \in \Z_{\geq 0}^S$.
Then, each reaction $r$ of order $\Order_r$ corresponds to exactly one monomial in $\mathbf{Y}^{\otimes \Order_r}$:
\begin{align*}
M_r(\mathbf{Y}) = \prod_{i=1}^S y_i^{\alpha_{r,i}} \leftrightarrow \text{position } \ell_r = \IndexFunc_{\Order_r}(\underbrace{1,\ldots,1}_{\alpha_{r,1}}, \underbrace{2,\ldots,2}_{\alpha_{r,2}}, \ldots, \underbrace{S,\ldots,S}_{\alpha_{r,S}}) \quad .
\end{align*}
Here the lexicographic index function $\IndexFunc_k: \{1,2,\ldots,S\}^k \to \{1,2,\ldots,S^k\}$ maps the ordered tuple $(i_1, i_2, \ldots, i_k)$ to its position in $\mathbf{Y}^{\otimes k}$:
\begin{align*}
    \IndexFunc_k(i_1, i_2, \ldots, i_k) = 1 + \sum_{j=1}^k (i_j - 1) S^{k-j} \quad ,
\end{align*}
which satisfies:
\begin{enumerate*}[label=(\roman*)]
    \item $\IndexFunc_k(1,1,\ldots,1) = 1$,
    \item $\IndexFunc_k(S,S,\ldots,S) = S^k$, and
    \item $\IndexFunc_k$ is bijective and order-preserving.
\end{enumerate*}
Mixed orders occur when a system contains multiple reactions with different orders. 
Examples of polynomial structure and reaction types are given in \Cref{sec:poly-structure}.  
According to \cref{eq:rd-stochi}, for a reaction network with reactions of order $\Order$, the coefficient tensor $\mathbf F_\Order \in \mathbb{R}^{S \times S^\Order}$ has the explicit form:
\begin{align}\label{eq:general_F_varsigma}
\mathbf{F}_\Order[i, \ell] = \sum_{r: \Order_r = \Order} (\beta_{r,i} - \alpha_{r,i}) c_r \delta_{\ell, \ell_r}
\end{align}
where $\ell_r$ is the tensor position corresponding to the reactant monomial of reaction $r$, and $\delta$ is the Kronecker-$\delta$. We give an example of autocatalytic reaction in \cref{sec:autocatalytic} and its explicit form of $\mathbf F_\varsigma$ in \cref{eq:autocatalytic_F_varsigma} because biological reproduction rests ultimately on chemical autocatalysis \citep{konnyu_kinetics_2024}.
With the explicit form of $\mathbf{F}_\varsigma$ in hand, we now bound its spectral norm that will be used for bounding the quantum complexity. 
For the general coefficient tensor $\mathbf{F}_\Order$, the $l_\infty$-norm satisfy:
\begin{align*}
\|\mathbf{F}_\Order\|_\infty = \max_i \sum_{\ell=1}^{S^\Order} |\mathbf{F}_\Order[i,\ell]| \leq \max_i \sum_{r=1}^R |(\beta_{r,i} - \alpha_{r,i})| c_r \leq C_{\max} \sigma_{\max}
\end{align*}
where $C_{\max} = \max_r c_r$ and $\sigma_{\max} = \max_i \sum_{r=1}^R |\beta_{r,i} - \alpha_{r,i}|$.
The $l_1$-norm satisfies 
\begin{align*}
\|\mathbf{F}_\Order\|_1 = \max_\ell \sum_{i=1}^S |\mathbf{F}_\Order[i,\ell]| \leq C_{\max} \sum_{i=1}^S \sum_{r: \ell_r = \ell} |\beta_{r,i} - \alpha_{r,i}|
\end{align*}
which leads to the bound of its $l_2$-norm:
\begin{align}
\|\mathbf{F}_\Order\|_2 \leq \sqrt{\|\mathbf{F}_\Order\|_\infty \cdot \|\mathbf{F}_\Order\|_1} \leq \sqrt{C_{\max} \sigma_{\max} \cdot C_{\max} \tau_{\max}} = C_{\max} \sqrt{\sigma_{\max} \tau_{\max}} \le C_{\max} s
\end{align}
where $\tau_{\max} = \sum_{i=1}^S \sum_{r: \ell_r = \ell} |\beta_{r,i} - \alpha_{r,i}|$ is the maximum total stoichiometric change per reaction position and $s$ is the sparsity of $\mathbf{F}_\Order$. The assumption of sparsity is valid and commonly made for reaction networks, as most species are unlikely to interact with all other species simultaneously \cite{mordvintsev2023differentiable}. 

\subsection{The canonical GM model}\label{sec:GM}

For the numerical simulations in \cref{sec: simulations}, we focus on a particular instance of the Turing model, the Gierer-Meinhardt (GM) model \cite{Gierer1972}, which has applications in pattern formation related to the regulation of the size and spacing of structures.  
The governing equation of the GM model is \citep{Gierer1972},
\begin{align}\label{eq:GM_a}
\frac{\partial y_1(x,t)}{\partial t} &= D_1 \nabla^2 y_1(x,t) - \mu_1 y_1(x,t) + c_1 b(x,t) y_1^2(x,t) y_2(x,t)+ b(x,t) b_1\\ 
 \label{eq:GM_s}
\frac{\partial y_2(x, t)}{\partial t} &= D_2 \nabla^2 y_2(x,t) -\mu_2 y_2(x,t) - c_1b(x,t) y_1^2(x,t) y_2(x,t) +b_2 , 
\end{align}
where $y_1$ and $y_2$ represents the evolution of the activator morphogen concentration and the inhibitor morphogen concentration, respectively. These equations are parameterized by diffusion coefficients $D_1$ and $D_2$ for the respective morphogen species, decay rate parameters $\mu_1$ and $\mu_2$, morphogen coupling parameter $b(x, t)$ \cite{gierer1972theory}, reaction rate coefficient $c_1$, and constant parameters $b_1$ and $b_2$. For a given reaction system, the decay rates and reaction coefficients can be determined through molecular dynamics simulations using thermodynamic integration \cite{Bhati2017} and transition state theory \cite{Vestergaard2016}. The term $y_1^2(x,t) y_2(x,t)$ in \cref{eq:GM_a} and \cref{eq:GM_s} describes an auto-induction process where an entity, in this case a chemical $y_1$, facilitates self-production consuming a chemical $y_2$ as a feedstock. 
Without loss of generality and with a focus on chemical reactions, we assume $b(x, t) = 1$, which simplifies \cref{eq:GM_a} and \cref{eq:GM_s} to
\begin{align}\label{eq:GM_a_rho1}
\frac{\partial y_1(x,t)}{\partial t} &= D_1 \nabla^2 y_1(x,t) - \mu_1 y_1(x,t) + c_1 y_1^2(x,t) y_2(x,t)+ b_1\\ 
 \label{eq:GM_s_rho1}
\frac{\partial y_2(x, t)}{\partial t} &= D_2 \nabla^2 y_2(x,t) -\mu_2 y_2(x,t) - c_1 y_1^2(x,t) y_2(x,t) +b_2. \quad
\end{align}

\subsection{Spatial Discretization of \cref{eq:Turing-poly}} \label{sec: Discrete} 
To analyze systems computationally, it is necessary to discretize continuous multi-dimensional space. 
Let $h$ be the spacing between each spatial nodes, $n$ be the number of nodes in each spatial dimension, and $n_d = n^d$ be the total number of nodes for $d$ spatial dimension.  
We implement the finite central difference method for approximating the Laplacian operator assuming periodic boundary conditions. The second derivative of a function \( f \) at a grid point \( x_j \) is approximated using the central difference formula:

\begin{equation*}
f''(x_j) \approx \frac{f(x_{j+1}) - 2f(x_j) + f(x_{j-1})}{h^2}.
\end{equation*}
This formula provides a second-order accurate approximation of the second derivative.

For a discretized grid, the Laplacian \(\Delta_h\) is constructed through Kronecker (or tensor) products. The Kronecker product is an operator on two matrices, which results in a block matrix. If $A = [a_{ij}] \in \mathbb{R}^{m\times n}$ and $B = [b_{kl}]\in \mathbb{R}^{p\times q}$, then the Kronecker product, denoted by $\otimes$,  $A \otimes B$ is given by:

\begin{align*}
A \otimes B = \begin{bmatrix}
a_{11}B & a_{12}B & \cdots & a_{1n}B \\
a_{21}B & a_{22}B & \cdots & a_{2n}B \\
\vdots & \vdots & \ddots & \vdots \\
a_{m1}B & a_{m2}B & \cdots & a_{mn}B \\
\end{bmatrix}
\end{align*}
where $A \otimes B \in \mathbb{R}^{mp\times nq}$. Using this notation, the Laplacian operator $\Delta_h$ is given by \citep{liu_efficient_2023}:
 
\begin{equation} \label{laplace0}
\Delta_h := \mathbf{\mathcal{D}_h} \otimes \underbrace{I \otimes \cdots \otimes I}_{d-1 \text{ terms}} + I \otimes \mathbf{\mathcal{D}_h} \otimes \underbrace{I \otimes \cdots \otimes I}_{d-2 \text{ terms}} + \cdots + \underbrace{I \otimes \cdots \otimes I}_{d-1 \text{ terms}} \otimes \mathbf{\mathcal{D}_h},
\end{equation}
where $h$ represents the distance between each node. The Laplacian operator \(\mathbf{\mathcal{D}_h}\) with central differences and periodic boundary conditions on a uniform grid for a one-dimensional grid of \(n\) points is given by
\begin{align} \label{laplace1}
    \mathbf{\mathcal{D}_h}=n^2\begin{bmatrix}
-2&1&0&\cdots&1\\1&-2&1&0&\cdots\\0&\ddots&\ddots&\ddots&0\\
        0&\cdots&1&-2&1\\1&\cdots&\cdots&1&-2
    \end{bmatrix}_{n\times n}.
\end{align}
Therefore the dissipative terms in RDEs can be rewritten as $D_i \nabla^2y_i=D_i\Delta_h\mathbf{\tilde{y}}_i $ for $i=1, 2, \cdots, S$. 
We refer to \cref{sec:laplacian-norm} for details of the spectral norm of \cref{laplace0} that is crucial for the quantum complexity analysis.
The spatially discretized form of the $i$-th instance of the $S$-species Turing model \cref{eq: Turing} is 
\begin{align}\label{eq: Discrete Turing}
\frac{\d\tilde{y}_i}{\d t} =\mathbf{\tilde{R}_i} \cdot \bigoplus_{j=0}^\varsigma \mathbf{\tilde{Y}}^{\otimes j} + D_i \Delta_h\tilde{y_i}, \quad \text{for } i=1, 2, \dots, S 
\end{align}
where $\tilde{y}_i = [y_{i_1}, y_{i_2}, \dots, y_{i_{n_d}}]^\top \in \mathbb{R}^{n_d}$ for $i\in [1, 2, \dots, S]$ is the spatially distretized solution vector of the $i$-th species, $\mathbf{\widetilde{Y}} = [\tilde{y}_1, \tilde{y}_2, \dots, \tilde{y}_S] \in \mathbb{R}^{n_d S}$ is the full solution vector, and $\mathbf{\tilde{R}_i}$ is the spatially discretised $\mathbf{R}_i$. The polynomial form of \cref{eq: Discrete Turing} is given as: 
\begin{align}\label{eq:discrete_Turing}
\frac{d\mathbf{\tilde{Y}}}{d t}=\mathbf{\tilde{F}}_0 + \mathbf{\tilde{F}}_1\mathbf{\tilde{Y}} + \mathbf{\tilde{F}}_2 \mathbf{\tilde{Y}}^{\otimes 2}+\dots +\mathbf{\tilde{F}}_\varsigma \mathbf{\tilde{Y}}^{\otimes \varsigma},
\end{align}
which is the discretized version of \cref{eq:Turing-poly}.

The spatially discretized form of $\mathbf F_1$ in \cref{eq:F1} is 
\begin{align}
\label{eq:F1-n}
\tilde{\mathbf{F}}_1 = 
\begin{bmatrix}
    \mathcal{D}_1 & 0 & 0 & 0 & \cdots & 0 \\
    0 & \mathcal{D}_2 & 0 & 0 & \cdots & 0\\
    \vdots & \vdots & \vdots & \vdots & \ddots \\
    0 & 0 & 0 & 0 & \cdots & \mathcal{D}_S\\
\end{bmatrix}
\in \mathbb{R}^{(S\cdot n_d)\times (S\cdot n_d)}
, 
\end{align}
where 
$\mathcal{D}_i = D_i \Delta h -\mu_i I$ 
for $i\in{1, 2, \dots, S}$ representing the coefficient of the total linear term. 
The spectral norm of diagonal linear coefficient matrix $\mathbf{\tilde{F}}_1$ in \cref{eq:F1-n} can be bounded by utilizing the norm of the discretized Laplacian operator (\cref{eq:laplacian-norm-1d}), 
\begin{equation}\label{eq:tildeF1-norm}
  \| \mathbf{\tilde{F}}_1 \|  \le \max_{i\in1, \dots, S}\| D_i \Delta h - \mu_i I \| \le 4 d n^2 \max_{i\in1, \dots, S} D_i + \max_{i\in1, \dots, S}|\mu_i| .
\end{equation}
For spatially discretized systems with \textit{local} reactions, the spatial coefficient tensor can be obtained by mapping \cref{eq:general_F_varsigma} via Kronecker-$\delta$:
\begin{align}\label{eq:spatial_tensor_formula}
\tilde{\mathbf{F}}_\Order[m, \ell] = \sum_{p=1}^{n_d} \sum_{i=1}^S \sum_{r \in \mathcal{R}_p} \delta_{m, (p-1)S + i} \cdot \delta_{\ell, \ell_{\text{global}}^{(p)}(r)} \cdot (\beta_{r,i} - \alpha_{r,i}) c_r^{(p)}
\end{align}
where $\mathcal{R}_p$ is the set of reactions at grid point $p$,
$c_r^{(p)}$ is the rate constant for reaction $r$ at point $p$, and
$\ell_{\text{global}}^{(p)}(r)$ is the global tensor position for reaction $r$ at point $p$.
The spectral norm of \cref{eq:spatial_tensor_formula} is then given by
\begin{align}\label{eq:tildeF-norm}
\|\tilde{\mathbf{F}}_\Order\|_2 = \max_{p} \|\mathbf{F}_\Order^{(p)}\|_2 \leq \max_{p} \mathcal C_{\max}^{(p)} \sqrt{\sigma_{\max} \tau_{\max}} \le \mathcal C_{\max} s
\end{align}
which is independent of $n_d$ due to properties of Kronecker-$\delta$.
For the autocatalytic case, \cref{eq:F-sigma-l2} leads to
\begin{equation}
\|\tilde{\mathbf{F}}_\Order\|_2 = \|\mathbf{F}_\Order\|_2 \leq \mathcal C_{\rm max} \sqrt{2(2S-1)}
\end{equation}
where $\mathcal C_{\rm max} = \max_{p=1, \dots, n_d} c_{ij}$.

\section{Results}
\label{sec:reaction-rate}
At first glance, the prospects for quantum advantage for these simulations may seem remote.  This is because any potential exponential advantage in scaling with respect to the size of the grid vanishes due to the Laplacian term diverging as $1/h^2$ in magnitude.  However, the reaction rate calculation provides an opportunity for substantial advantage with respect to classical methods.  This is because the reaction rates are directly related to the free energy differences for a reaction and these differences are believed to be exponentially hard to estimate using a classical computer at low temperature.  The Eyring equation \citep{eyring_activated_1935} states that the reaction rate between two configurations, $i,j$ along the reaction coordinates is given by 

\begin{equation}
    c_{i,j} = \frac{k_BT}{2\pi}e^{-\Delta G/k_BT}, 
\end{equation}
where $k_B$ is Boltzmann's constant, $T$ is the temperate, and Gibbs free energy, $\Delta G = \Delta H - T\Delta S$, is the free-energy difference between the two coordinates and we use units where $\hbar=1$ above.  If $\Delta G <0$ then the reaction is spontaneous in the forward direction that form products, if $\Delta G >0$ then it will spontaneously proceed in the reverse direction and favor the formation of reactants, and if $\Delta G=0$ then the system is in equilibrium thus no net change in concentrations of the products and reactants. That means that the reaction rate can be computed from the difference in enthalpy and entropy between two different configurations.  The entropy difference between two different configurations is related to the change in the number of degrees of freedom that a system experiences in a reaction and can often be estimated using classical computers~\cite{reiher2017elucidating}.  For strongly correlated quantum systems, the enthalpy difference is conjectured to be exponentially hard to compute classically~\cite{o2022intractability}, but can be polynomially expensive on a quantum computer if a state of inverse polynomial overlap can be prepared.

The complexity of the enthalpy computation on a quantum computer can be computed in polynomial time provided that several criteria are met.  For simplicity we focus on second quantization in the following, but first quantization can also have a myriad of advantages in terms of the scaling of the Hilbert-space dimension~\cite{su2021fault,kivlichan2017bounding}.

\begin{enumerate}
\item The system can be accurately discretized within error $\epsilon$ to a number of orthogonal spin orbitals $N$ that is at most inverse polynomial in the desired error tolerance $\epsilon$.
    \item The matrix elements of the Coulomb Hamiltonian can be computed in time polynomial in $N$ within the above basis.
    \item There exists a state $\ket{\psi}\in\mathbb{C}^{2^N}$ that can be prepared in time that is polynomial in $N$ on a quantum computer and there exists a known $\delta>0$ such that $|\braket{\psi|\psi_0}|^2 \ge 1-\delta^2$ for groundstate (or more generally target state) $\ket{\psi_0}$ within the basis.
    \item A constant is known $\Delta>0$ and an energy estimate for the groundstate energy $\hat{E}_0$ such that the true excited state energy satisfies $E_1 \ge E_0 +\Delta$.
\end{enumerate}

Under these circumstances, the number of times a subroutine computing the matrix elements of the Hamiltonian needs to be queried to prepare such a state within error $O(\epsilon)$ is~\cite{lin2022heisenberg}
\begin{equation}
    O\left(\frac{{\rm polylog}{(1/\epsilon)}}{\Delta\delta} \right).
\end{equation}
In turn under certain assumptions about the scaling of the rank for state-of-the-art double factorization/hypercontraction approaches, the number of gate operations needed in this has been shown to be in~\cite{lee2021even}
\begin{equation}
    O\left(\frac{{N^2{\rm polylog}(N/\epsilon)}}{\Delta\delta} \right).
\end{equation}
An important distinction that arises here is that the cost of estimating the groundstate energy as a bit string is actually far more expensive than this.  However, for our application we do not need to actually estimate the groundstate energy.  Rather we will see that we only actually need to build a subroutine that can return the groundstate energy differences, and in turn the reaction rates, within a block encoding.  We do this because the cost of such an estimate is 
\begin{equation}
    O\left(\frac{N^2{\rm polylog}{(1/\epsilon)}}{\epsilon\Delta\delta} \right),
\end{equation}
although $\sqrt{\epsilon}$ scaling can be achieved if promises are made about the structure of the decomposition~\cite{simon2024amplified,king2025quantum}.  In practice, the precision needed for a single reaction rate calculation is large: an error of roughly $2$ mHa is needed in these calculations.  Thus if we have a multitude of different states to consider, the costs of classically learning these reaction rates will often be prohibitive.

Entropy calculation on a quantum computer can be a difficult problem.  The issue is that the entropy is ${\rm Tr}(-\rho \log(\rho))$.  The von Neumann entropy is difficult to estimate because of the non-analyticity of the entropy for eigenvalues of $\rho$ near zero.  The complexity of estimating the von Neumann entropy within constant multiplicative error scales polynomially with the dimension of $\rho$~\cite{gur2021sublinear,gur2021sublinear}.  This implies that a polynomial time algorithm for the entropy in full generality is impossible in the case of multiplicative error.  Additive error is more positive, however, and the error of entropy estimation for high-temperature thermal distributions can be efficiently estimated for high-temperature thermal distributions~\cite{simon2024improved} using purely quantum mechanical approaches.

A simple approximate approach that can be taken for this involves the use of the Zwanzig equation which states that
\begin{equation}
    \Delta G \approx -k_BT \ln\left( \left \langle e^{-(E_A - E_B)/k_BT} \right\rangle_A\right)
\end{equation}
which allows us to find an estimate of the free energy difference from the energy where the average is taken over Metropolis-Hastings trajectories over the set of different possible reaction coordinates that can be considered where $A$ is sampled from the Gibbs distribution $e^{-H_i/k_BT}/Z_i$ and $B$ is an energy sampled to $e^{-H_j /k_BT}/Z_j$ assuming that the process is isothermal.  

This process is tractable on a quantum computer, but requires sampling.  The sampling process is complicated and so a common approach that can be used is to consider a perturbation result that only looks at small differences over the reaction coordinates.  Specifically, the second order expansion takes the form~\cite{chipot2007free}
\begin{align}
    \Delta G &\approx  \langle H_j - H_i\rangle_i -\frac{1}{2k_BT}\left(\langle (H_j - H_i)^2\rangle_i -\langle H_j - H_i \rangle^2_i  \right)\nonumber\\
    &= \langle H_j - H_i\rangle_i - \frac{1}{2k_BT}\left(\langle H_i^2\rangle_i  + \langle H_j^2\rangle_i - \langle H_i\rangle^2_i - \langle H_j\rangle^2_i -2{\rm Re}\langle H_i H_j \rangle_i \right)
\end{align}
where here $\langle \cdot \rangle_i$ refers to ${\rm Tr}((\cdot)e^{-H_i/k_BT}/Z_i)$. 

If we assume that the gap of $H_i$ from the groundstate is large with respect to $k_BT$ which is in turn large with respect to the groundstate energy then we can treat the unperturbed state of $H_i$ to be the eigenstate $\ket{\psi_i}$.  In this case the free energy difference reads

\begin{align}\label{eq:DeltaG}
    \Delta G &\approx \bra{\psi_i} H_j \ket{\psi_i} - E_i - \frac{1}{2k_BT}\left( \bra{\psi_i} H_j^2\ket{\psi_i}  - \bra{\psi_i} H_j\ket{\psi_i}^2 -2{\rm Re}\bra{\psi_i} H_i H_j \ket{\psi_i} \right)
\end{align}

We will now discuss the complexity of block-encoding the Zwanzig approximation to the Free-Energy difference using a truncated Taylor series approximation under the above assumptions.  For clarity we will begin by discussing how to block encode the various components in the $\Delta G$ expansion.

\begin{lemma}\label{lem:stateBE}
    Let $H_i = \sum_j \alpha_j(i) U_j$ that is block-encoded  $\ket{\phi_i}$ such that $|\bra{\psi_1}\phi_1\rangle|^2\ge 1-\delta^2$ with the promise that a value $E_{0,i}$ is known such that all eigenvalues other than the groundstate are promised to be larger than $E_{0,i} + \gamma$ and are lower than $E_{0,i}-\gamma$ for a positive constant $\gamma$. There exists a quantum algorithm that yields an $\epsilon$-approximate block encoding of $\ket{\psi_i}$, $V_{\psi_i}$, such that
    $$
    (\bra{0}\bra{0}\otimes  I) V_{\psi_i} (\ket{0}\ket{0}\otimes I) = \ket{\psi_i}\!\bra{0} + O(\epsilon)
    $$
    that uses a number of queries to $U_{\psi_i}$ that scales as $\widetilde{O}(\sum_j |\alpha_j(i)|\log(1/\epsilon)/\gamma \delta)$. Similarly the cost of block encoding $\bra{\psi_i}$ is also given by the above result.
\end{lemma}
\begin{proof}
    Using the results of~\cite{Gilyen2019} we can use fixed point amplitude amplification in order to construct a unitary $V$ such that   \begin{equation}
       (\bra{0}\bra{0}\otimes I)V \ket{0}\ket{0}\ket{0} = \ket{\psi_i} + O(\epsilon)
   \end{equation} 
    the groundstate within error $\epsilon$ using a number of queries to $U_{\psi_i}$ that is in
    \begin{equation}
        C(V):=\widetilde{O}(\sum_j |\alpha_j(i)|\log(1/\epsilon)/\gamma\delta)\label{eq:CV}
    \end{equation}
    This does not, however imply that this unitary is equivalent to $V_{\psi_i}$ because the $V_{\psi_i}$ operation needs to zero out the off-diagonal matrix elements within the block.  However, since we do not know the action of the unitary on the remaining components it is difficult to yield the matrix.  We have however that
    \begin{equation}
        V(\ket{0}\!\bra{0}) = \ketbra{\psi_i}{0}
    \end{equation}
    as required.  This implies that we can write
    \begin{equation}
        \frac{V}{2}s(I+(I -2 \sum_{j\ne 0} \ket{j}\!\bra{j})) = \ketbra{\psi_i}{0}.
    \end{equation}
    This is a linear combination of two unitaries with a block-encoding constant of $1$.  Specifically we have that
    \begin{equation}
        (H\otimes I)(\ketbra{0}{0}\otimes V  + \ketbra{1}{1}\otimes V(I -2 \sum_{j\ne 0} \ket{j}\!\bra{j})(H\otimes I)
    \end{equation}
    provides a $(1,1,\epsilon)$ block encoding of $\ketbra{\psi_0}{0}$ using $\widetilde{O}(\sum_j |\alpha_j(i)|\log(1/\epsilon)/\gamma\delta)$ queries to $U_{\psi_i}$ from~\eqref{eq:CV}.  As the adjoint of the block encoding is the block encoding of the adjoint, we can use the same procedure to block encode the adjoint at the aforementioned cost.
\end{proof}

\begin{lemma}\label{lem:BE}
    Let $k$ index an element in a finite set of Hamiltonians such that $H_k = \sum_{j} \alpha_j(k) U_j$ and assume that there exist for each $k$ an operator ${\rm PREP}_k$ such that ${\rm PREP}_k \ket{0} = \sum_{j} \sqrt{\alpha_j(k)} \ket{j}/\sqrt{\alpha(k)}$ and ${\rm SELECT} = \sum_j \ketbra{j}{j}\otimes U_j$ for unitary $U_j$.  A zero-error block encoding of $H_k$ can be implemented with a normalization constant of $\alpha(k)$ using at most $3$ queries to ${\rm PREP}_k,{\rm SELECT}$ or their adjoints.
\end{lemma}
\begin{proof}
    The proof immediately follows from the LCU lemma~\cite{childs2012hamiltonian}.
\end{proof}

\begin{lemma}\label{lem:DeltaGBd}
For any $k_BT>0$ an $\epsilon$-approximate block encoding of the constant  approximation to  $\Delta G$ can be constructed using a number of queries to the ${\rm PREP}_k$ and {\rm SELECT} operations that is in 
$$
\widetilde{O}\left(\frac{\max_j\alpha(j)(1+\alpha(i)/k_BT)\log(1/\epsilon)}{\gamma\delta}\right)
$$
with a block encoding constant of $O(\max_j \alpha(j)(1 + \alpha(i)/k_BT))$
\end{lemma}
\begin{proof}
From~\cite{gilyen_quantum_2019} we have that a block encoding of a sum of two operators can be formed with zero error using a block encoding constant that is the sum of the block encoding constants of the two unitaries.  Similarly, the work also shows that a block encoding of the product of two matrices can be formed with zero error using two applications of the block encoding matrices with zero error and the product of the two block encoding constants.

We can then provide block encodings of the constants in the approximation to $\Delta G$ in terms of the following operators.  Let $V_{1}$ be a unitary such that
\begin{equation}
    (\bra{0}\otimes I) V_{1} (\ket{0}\otimes I) = \frac{\bra{\psi_i} H_j  \ket{\psi_i} +O(\epsilon_1)}{\beta_j}\ketbra{0}{0}
\end{equation}
Similarly let $V_{2},\ldots,V_{5}$ be unitaries such that
\begin{align}
    (\bra{0}\otimes I) V_{2} (\ket{0}\otimes I) &= \frac{\bra{\psi_i} H_i  \ket{\psi_i} +O(\epsilon_2)}{\beta_2}\ketbra{0}{0} \nonumber\\
    (\bra{0}\otimes I) V_{3} (\ket{0}\otimes I) &= \frac{\bra{\psi_i} H_j H_j  \ket{\psi_i} +O(\epsilon_3)}{\beta_3}\ketbra{0}{0} \nonumber\\(\bra{0}\otimes I) V_{4} (\ket{0}\otimes I) &= \frac{\bra{\psi_i} H_i  \ket{\psi_i}\bra{\psi_i} H_i  \ket{\psi_i} +O(\epsilon_4)}{\beta_4}\ketbra{0}{0} \nonumber\\
    (\bra{0}\otimes I) V_{5} (\ket{0}\otimes I) &= \frac{\bra{\psi_i} H_i  \ket{\psi_i}\bra{\psi_i} H_j  \ket{\psi_i} +O(\epsilon_5)}{\beta_5}\ketbra{0}{0} \nonumber\\
\end{align}

Using the product properties of block encodings we have from \cref{lem:stateBE} that $\ket{\psi_i}$ can be block-encoded with constant $1$ within error $\epsilon'$ using $O(\alpha(i)\log(1/\epsilon')/\gamma \delta)$ queries.  The same result also holds for a block-encoding of $\bra{\psi_i}$.  \cref{lem:BE} shows that $H_j$ can be block encoded with constant $\alpha(j)$ using $O(1)$ queries.  Thus by the product property of block-encodings we have that we can block-encode the product of these three operations such that
\begin{equation}
    \frac{\bra{\psi_i} H_j  \ket{\psi_i} +O(\epsilon_1)}{\beta_j}\ketbra{0}{0} = \frac{\bra{\psi_i} H_j  \ket{\psi_i} +O(\epsilon' \alpha(j))}{\alpha(j)}\ketbra{0}{0}
\end{equation}
The same reasoning can be used to show that $E_i \ketbra{0}{0}+O(\epsilon_2)$ can be block encoded using $O(\alpha(i) \log(1/\epsilon_2)/\gamma \delta)$ queries such that 
\begin{equation}
    \beta_2 = \alpha(i). 
\end{equation}

The higher-order block encodings can be similarly produced by constructing the products of block encodings.  We have that $\bra{\psi_i} H_j H_j \ket{\psi_i}$ can be written as the product of $4$ block encodings.  The block-encoding constant of this product is again from~\cite{gilyen_quantum_2019} to be
\begin{equation}
    \beta_3 = 1\cdot \alpha(j) \cdot \alpha(j)\cdot 1= \alpha^2(j).
\end{equation}
The error in the block encoding is $O(\alpha_j \epsilon')$ where $\epsilon'$ is the error in the block-encoding of $\ket{\psi_i}$ from~\cite{gilyen_quantum_2019}.

The exact same reasoning can be used to show that $\beta_4 = \alpha(i)^2$ and $\beta_5 = \alpha(i) \alpha(j)$ and the error in the block encodings is $\max(\epsilon_4,\epsilon_5)=O(\max_j \alpha(j) \epsilon')$.  In each case the block encoding of the Hamiltonian terms can be implemented using $O(1)$ queries and the four remaining state encodings can be implemented using $O(\alpha(i) \log(1/\epsilon')/\gamma\delta)$ queries to PREP and SELECT oracles.

The approximation to $\Delta G$ can then be formed by summing these block encodings using linear combination of unitaries.  Specifically we can construct an operator $W$ such that
\begin{equation}
    (\bra{0}\otimes I) W (\ket{0}\otimes I) = \frac{\Delta G + O(\epsilon)}{\alpha}
\end{equation}
The block encoding constant is the sum of the block encoding constants and in turn
\begin{equation}
    \alpha = (\beta_1+\beta_2) +\frac{\beta_3 +\beta_4 +\beta_5}{k_BT} = O\left( \max_j \alpha(j)\left( 1+ \frac{\max_j\alpha(j)}{k_BT} \right) \right)
\end{equation}
we can take $\epsilon'= O(\epsilon/(1+\max_j \alpha(j)/k_BT))$ which leads to a query complexity for the block encoding of
\begin{equation}
    \widetilde{O}\left(\frac{\max_j\alpha(j)(1+\alpha(i)/k_BT)\log(1/\epsilon)}{\gamma\delta}\right).
\end{equation}
\end{proof}
Now that we know how to construct a block encoding of the approximation to the free energy difference given in~\eqref{eq:DeltaG}.  The next step is to provide a block encoding of the reaction rate.
\begin{theorem}\label{thm:UC}
    A unitary operator $U_C$ can be constructed such that
    $$
    (\bra{0}\bra{i}\bra{j} ) U_C (\ket{0}\ket{i}\ket{j} ) = \frac{e^{-\Delta G(ij)/k_BT} +O(\epsilon)}{\alpha_{\exp}}. 
    $$
    using a number of queries to ${\rm PREP}_k$ and {\rm SELECT} that are in
    $$
    \widetilde{O}\left(  \frac{\max_j\alpha(j)(1+\max_i\alpha(i)/k_BT)\log^2(1/\epsilon)}{\gamma\delta}\right)
    $$
    for 
    $$
    \alpha_{\exp} \in e^{ O(\frac{\max_j \alpha(j)}{k_B T}(1+\max_j\alpha(j)/k_{B}T)    } 
    $$
\end{theorem}
\begin{proof}
From Lemma~\ref{lem:DeltaGBd} we have that we can construct for any $i,j$ a block encoding of $\Delta G(ij) \ketbra{0}{0}$, the approximate free energy difference between the two configurations, which we denote $U_{\Delta G(ij)}$ using a number of queries that are in 
\begin{equation}
    \widetilde{O}\left(\frac{\max_j\alpha(j)(1+\alpha(i)/k_BT)\log(1/\epsilon)}{\gamma\delta}\right)
\end{equation}
If we replace the PREP operations with the more specific ${\rm PREP}(ij)$ (the SELECT can be kept the same by assumption) we have that a block encoding of $\sum_j \ketbra{ij}{ij}\otimes \Delta G(ij)\ketbra{0}{0}$ through the following steps.
\begin{enumerate}
    \item Choose PREP$_0=\bigotimes_{p=0}^{M^2-1} H$
    \item Combine this PREP with the controlled PREP via PREP$_1=(\bigotimes_{p=0}^{M^2-1} H)(\sum_{i,j=0}^{M-1} \ketbra{ij}{ij}\otimes{\rm PREP}(ij))$.
    \item Use $U_{\Delta G}={\rm PREP}_1^\dagger (I\otimes {\rm SELECT}) {\rm PREP}_1 $ as our block encoding unitary.
\end{enumerate}
The number of queries needed to produce our unitary $U_{\Delta G}$ is then
\begin{equation}
    \widetilde{O}\left(\frac{\max_j\alpha(j)(1+\max_i\alpha(i)/k_BT)\log(1/\epsilon)}{\gamma\delta}\right)
\end{equation}
Then from the LCU lemma~\cite{childs2012hamiltonian} this can be used to block encode $\Delta G(ij)$ using a block encoding constant equal to the maximum of those needed for each of the blocks.  This is 
\begin{equation}
\alpha_{\Delta G} \in O(\max_j \alpha(j)(1+\max_j\alpha(j)/k_{B}T)    
\end{equation}
from \cref{lem:DeltaGBd}.  Similarly the error in this block encoding is at most the maximum error in each of the individual control blocks.  Thus no error propagation is needed from the initial error in the block encoding of the $\Delta G(ij)$ values, which we denote $\epsilon_0$

 Next we need to construct the exponential of $\Delta G(ij)$.  A standard approach to perform this block encoding would involve using a Taylor series expansion of the exponential which we take for simplicity.  We take an approximation of the form
 \begin{equation}
    \|\sum_{j=0}^K (-\Delta G(ij)/k_BT)^j/j! - e^{-\Delta G(ij)/k_BT}\|\le \epsilon
 \end{equation}
From Taylor's theorem we have that the remainder term, after using a standard argument involving Stirling's inequality and asymptotic bounds on the LambertW function, leads to
\begin{equation}
    K \in O\left(\frac{\log(\max \Delta G(ij)/k_BT)}{\log\log(\max \Delta G(ij)/k_BT)} \right).
\end{equation}
Thus we can perform a linear combination of the block encodings of $\Delta G$ within error $\epsilon$ using $K$ multiplications of block encodings.  The block-encoding constants are at most the products of sums of these block encoding constants in this case.  This gives us
\begin{equation}
    \alpha_{\exp} \le e^{\alpha_{\Delta G}/k_BT}
\end{equation}
and requires $K$ queries to $U_{\Delta G}$ using the techniques of~\cite{berry2015simulating}.  The error in the block encoding is on the order of $K\alpha_{\Delta G}\epsilon_0$.  Now if we take $\epsilon_0 = \Theta(\epsilon/\alpha_{\Delta G})$ then we have that the error obeys
\begin{equation}
    K\alpha_{\Delta G} \epsilon_0 \in \widetilde{O}(\epsilon).
\end{equation}
Using a number of queries that scale as
\begin{equation} \label{eq:totalQueries}
    K N_{\rm queries}(\Delta G) = \widetilde{O}\left(  \frac{\max_j\alpha(j)(1+\max_i\alpha(i)/k_BT)\log^2(1/\epsilon)}{\gamma\delta}\right)
\end{equation}
The final claims then follow by substitution into the above results.
\end{proof}

\begin{corollary}\label{cor:alpha-F-varsigma}
Under the assumptions of \cref{thm:UC} we have that the matrix $\tilde{\mathbf{F}}_j$ can be block encoded within error $\epsilon$ specified in the Euclidean norm using a number of queries to ${\rm PREP}_k$, {\rm SELECT} and a new stochiometric oracle $U_{stoc}$ such that
$$
U_{stoc}\ket{0}= \sum_{ij} \sqrt{|\alpha(ij)-\beta(ij)|}\ket{ij}\ket{\delta_{\alpha(ij)<\beta(ij)}}
$$
that scales at most as  
\begin{equation}\label{eq:alpha-reactionRate}
\widetilde{O}\left(  \frac{\max_j\alpha(j)(1+\max_i\alpha(i)/k_BT)\log(\alpha_{\Delta G}/k_BT)\log^2(\sum_{ij}|\alpha(ij)-\beta(ij)|/\epsilon)}{\gamma\delta}\right)
\end{equation}
with a block encoding constant of 
\begin{equation}\label{eq:alpha-F-varsigma}
\alpha_F:=\sum_{ij} |\alpha(ij) - \beta(ij)| \alpha_{\exp} \in  O\left(\sum_{ij}|\alpha(ij) - \beta(ij)| e^{\frac{\max_j \alpha(j)}{k_B T}(1+\max_j\alpha(j)/k_{B}T)    } \right).
\end{equation}
Recall that $\alpha$ and $\beta$ represent block-encoding constant for stoichiometric coefficients defined right below \cref{eq:general_reaction}. 
\end{corollary}
\begin{proof}
    First note that from \cref{thm:UC} we can construct a block encoding of the exponential of the free energy differences.  The tensor $\tilde{\mathbf{F}}_j$ can  be block encoded in the following steps
    \begin{enumerate}
        \item Prepare the state $\ket{0}$
        \item Apply the transformation $\ket{0} \mapsto \ket{\alpha_{stoc}}:=\sum_{ij} \sqrt{|\alpha(ij) - \beta(ij)|}\ket{ij}\ket{\delta{\alpha(ij)<\beta(ij)}}/\sqrt{\alpha_{stoc}}$ using a query to the stochiometry oracle $U_{stoc}$.
        \item Apply $U_C\otimes I$ to $\ket{0}\ket{\alpha_{stoc}}$ for temperature $T'=2T$.
    \end{enumerate}
    This process implements from \cref{thm:UC} a unitary $U_{PREP}'$ such that
    \begin{equation}
        U_{PREP}' \ket{0} = \sum_{ij}\ket{0} \frac{\sqrt{|\alpha(ij) - \beta(ij)|}}{\sqrt{\alpha_{stoc}}} \frac{\sqrt{c_{ij}}}{\sqrt{\alpha_{\exp}}}\ket{ij}
    \end{equation}
    Now we can construct a block encoding of the operation $\ketbra{j}{i}$ by block encoding $Z\otimes\ketbra{j}{j} U_{+}^{i-j} = (1-(1-2\ketbra{j}{j}))U_{+}^{i-j}/2$ where $U_+$ is the unitary incrementer operation and the Pauli $Z$ operator acts on the register containing $\ket{\delta_{\alpha(ij)<\beta(ij)}}$ which serves to flip the sign of the matrix elements based on the sign of the stochiometric constants to undo the absolute values used in the prepare operations.  This process can be implemented using adders and a maximally controlled $Z$-gate, a polynomial number of single-qubit $X$ gates and  a single query to $U_C$.  Thus from the LCU lemma, the required block encoding can be built within error $\epsilon$ using the number of queries  in \eqref{eq:totalQueries} with a normalization constant of $\alpha_{stoc} \alpha_{\exp}$.  Note that because the error of the product of block encodings scales as the product of the error and the normalization factors~\cite{gilyen_quantum_2019} the final error 
    \begin{equation}\label{eq:alpha-F-error}
    \epsilon\rightarrow \epsilon/\sum_{ij}|\alpha(ij) - \beta(ij)|.
    \end{equation}
\end{proof}

\subsection{Carleman Embedding for Coupled 
Systems}  

To simulate biological reaction diffusion systems on a quantum computer, we need to map their nonlinear governing equations, i.e.,  \cref{eq: Discrete Turing}, to linear equations due to the linearity of quantum mechanics.
One of the widely used mapping techniques is Carleman linearization, which has been shown to be efficient in linearizing nonlinear fluid dynamics \citep{liu_efficient_2021, li_potential_2025} and other systems \citep{liu_efficient_2023} for applying quantum computing. Carleman linearization maps a system of nonlinear differential to an infinite-dimensional system of
linear differential equations \citep{Carleman1932, forets_explicit_2017, liu_efficient_2021}.
This is done by introducing auxiliary variables to represent nonlinear terms. The nonlinear terms are then expanded into an infinite series. A transfer matrix is constructed by arranging the expanded nonlinear terms and coefficients into a matrix where each row corresponds to a different order of the expansion and each column corresponds to a different auxiliary variable. The infinite linear system can be truncated into a finite system and solved. This gives a linear differential equation that is potentially solvable using quantum algorithms. Carleman linearization constructs a finite-dimensional approximation of the Koopman operator by expanding the dynamics in polynomial observables \cite{mauroy2020koopman}. 

We follow Carleman embedding method by \citet{Kowalski1991, Ude2001, forets_explicit_2017, liu_efficient_2023} but with a slight modification such that we do not include the inhomogenous terms inside the Carleman transfer matrix. Instead, we only add the inhomogeneous terms to the linearized homogeneous nonlinear PDEs. This method can simplify calculating the spectral norm of the coefficient matrix of Carleman-linearized systems for quantum query complexity. Proof of convergence for this method is an open question, though numerical simulations presented in \cref{sec: simulations} suggests rapid convergence of truncation error. 
Specifically, we Carleman-linearize \cref{eq:discrete_Turing} without the inhomogeneous term $\tilde{\mathbf{F}}_0$, which will be dealt with later. The Carleman linearization procedure recast \cref{eq:discrete_Turing} into linear equations of newly introduced Carleman variables $\mathbf{Z}_i$ that represents arbitrary polynominal orders of $\tilde{\mathbf{Y}}$, i.e.,  $\mathbf{Z}_i = \tilde{\mathbf{Y}}^{\otimes i}$ for $i \in \mathbb{N}$ ($\mathbb{N}$ represents the natural number). The dynamics of $\mathbf{Z}_i$ is determined by \cref{eq:discrete_Turing} following
the Leibniz product rule \citep{forets_explicit_2017}:
\begin{align}
\frac{d\mathbf{\tilde{Y}}^{\otimes i}}{dt}=\sum_{v=1}^{i} (\mathbf{\tilde{Y}} \otimes \cdots \otimes \underbrace{\sum_{j=0}^{\varsigma}\mathbf{\tilde{F}}_i\mathbf{\tilde{Y}}^{\otimes j}}_{v\text{-th place}}\otimes \cdots \otimes \mathbf{\tilde{Y}}) \quad , 
\end{align}
which can be written as
\begin{align}\label{eq:tildeY-tensor}
\frac{d\mathbf{\tilde{Y}}^{\otimes i}}{d\tilde{t}}=\sum_{j=0}^{\varsigma}\mathbf{B}_j^i\mathbf{\tilde{Y}}^{\otimes (j+i-1)} \quad, 
\end{align}
or
\begin{align}\label{eq:Zi}
    \frac{d\mathbf{Z_i}}{dt}=\sum_{j=0}^n \mathbf{B}_j^i\mathbf{Z_{j+i-1}} , \quad i\in \mathbb{N}
\end{align}
by introducing the linear mapping transfer matrix using differentiating and applying Leibniz rule:
\begin{align}\label{eq: liebniz}
\mathbf{B}^i_{j} &=\sum_{v=1}^{i}(\mathbb{I} \otimes \cdots \otimes \mathbb{I} \otimes {\mathbf{\mathbf{\tilde{F}}_i(t)}}\otimes
\mathbb{I} \otimes \cdots \otimes \mathbb{I}) \\
&=\mathbf{B}_j^1\otimes \mathbb{I}^{\otimes (i-1)}+ \mathbb{I}\otimes \mathbf{B}_j^{i-1},
\end{align}
where $\mathbf{\mathbf{\tilde{F}}}_i(t)$ appears in $v\text{-th position in the }i\text{-th fold Kronecker product}$ and $\mathbb{I}$ is the identity matrix. 
The indices $i$ and $j$ represents the row and  the column by incrementing starting from zero after the padding zero matrices in the row, respectively. $\mathbf{B}_0^i$ is always the element on the diagonal. There are always $i-1$ matrices of leading zeros in the $i^{th}$ row \footnote{Another standard notation is where $i$ is the row and $j$ is the column, though we follow the notation of \citet{Kowalski1991}. A quick conversion to this notation would be the $B$ lower number is $i+j$.}. 
The matrix form of \cref{eq:Zi}, excluding the inhomogeneous terms, is
\begin{align}\label{eq:dZM}
\frac{d}{dt}
\begin{bmatrix}
    \mathbf{Z}_1 \\ \mathbf{Z}_2 \\ \vdots \\ \mathbf{Z}_\varsigma \\ \vdots \\ \vdots
\end{bmatrix}
=
\begin{bmatrix}
    \mathbf{B}_0^1 & \mathbf{B}_1^1 & \cdots & \mathbf{B}^1_\varsigma& \mathbf{0} & \cdots &\cdots \\
    \mathbf{0} & \mathbf{B}_0^2 & \mathbf{B}_1^2 & \cdots & \mathbf{B}_\varsigma^2  & \cdots &\cdots\\
    \mathbf{0} & \mathbf{0} & \mathbf{B}_0^3 & \mathbf{B}_1^3 & \cdots &\mathbf{B}_\varsigma^3 & \cdots  \\
    \vdots & \vdots & \vdots & \vdots & \ddots & \cdots &\cdots \\
    \vdots & \vdots & \vdots & \vdots & \ddots & \cdots &\cdots \\
    \vdots & \vdots & \vdots & \vdots & \ddots & \cdots &\cdots
\end{bmatrix}
\begin{bmatrix}
    \mathbf{Z}_1 \\ \mathbf{Z}_2 \\ \vdots \\ \mathbf{Z}_\varsigma \\ \vdots \\ \vdots
\end{bmatrix}
\end{align}
The compact form of \cref{eq:dZM} is
\begin{align}\label{eq:Carleman form}
\frac{d\mathbf{Z}}{dt}=\mathbf{M}_\infty \mathbf{Z} + \mathcal{B}, 
\end{align}
where
\begin{align}
\mathbf{Z} &=
\begin{bmatrix}
\mathbf{\tilde{Y}^{\otimes1}} & \mathbf{\tilde{Y}^{\otimes2}} & 
\cdots &
\mathbf{\tilde{Y}^{\otimes \varsigma} \cdots}
\end{bmatrix}^\top \\
& = 
\begin{bmatrix}
\mathbf{Z}_1 & \mathbf{Z}_2 & 
\cdots &
\mathbf{Z}_\varsigma \cdots
\end{bmatrix}^\top
\end{align}
and $\mathbf{M}_\infty$ is an infinite-dimensional block upper-triangular matrix.
Here, we reintroduce the inhomogeneous term $\mathcal{B}$ back to \cref{eq:dZM}, 
\begin{equation}
  \mathcal{B}=\tilde{\mathbf{F}}_0 \oplus \mathbf{0} 
  .
\end{equation}
We then truncate the infinite Carleman linearized system (\cref{eq:Carleman form}) to $k$-th truncation order to obtain a finite linear ODE system,
\begin{align}\label{eq:Zk}
\frac{d\mathbf{Z_k}}{dt}=\mathbf{M_k\mathbf{Z_k}} + \mathcal{B}_k.
\end{align}
For example for $k=3$, 
\begin{align*}
\mathbf{M}_3=\begin{bmatrix}
        \mathbf{B}_0^1 & \mathbf{B}_1^1 & \mathbf{B}_2^1\\
        \mathbf{0} & \mathbf{B}_0^2 & \mathbf{B}_1^2 \\
        \mathbf{0} & \mathbf{0} & \mathbf{B}_0^3 
    \end{bmatrix}
\end{align*}
We apply the Carleman linearization procedure to the GM model, as shown in more detail in \cref{sec:example}, for numerical simulations presented in \cref{sec: simulations}. 
This method may provide quicker convergence and minimize error, as the inhomogeneous term is exact rather than an approximation. The original Carleman method may also be used for this workflow, and convergence has already been proved by \citet{liu_efficient_2023} for a non-coupled RDE with mixed boundary conditions. This proof can be modified to describe the coupled system with periodic boundary conditions. 

\subsection{Quantum Algorithm for RDEs}

The solution of the Carleman-linearized RDE, \cref{eq:Zk}, can be represented as
\begin{equation}\label{eqn:cre_solu}
  \mathbf{Z}_k(t) = \mathcal{T}e^{-\int_0^t \mathbf{M}_k \ud s} \mathbf{Z}_k(0) + \int_0^t \mathcal{T}e^{-\int_s^t \mathbf{M}_k \ud s'} \mathcal{B} \ud s,
\end{equation}
where $\mathcal{T}$ denotes the time-ordering operator. 
Quantum algorithms for simulating \cref{eq:Zk} aim at preparing an $\epsilon$-approximation of the quantum state $\ket{\mathbf{Z}(t)} = \mathbf{Z}(t)/\|\mathbf{Z}(t)\|$ with $t$ denoting the final evolution time. One of the most natural ways to solve linear PDEs is to map them into Schr\"odinger equation that are readily solvable on a quantum computer. We use linear combination of Hamiltonian simulations (LCHS) for this mapping as LCHS has been proclaimed to achieve both optimal state preparation cost and near-optimal scaling in matrix queries on all parameters \citep{an_linear_2023, an2023quantum}. Recall that both the coefficient matrix $\mathbf{M}_k$ and the inhomogeneous term $\mathcal B$ of \cref{eq:Zk} are time-independent, we therefore adopt the time-independent LCHS of \citet{an2023quantum}. The LCHS starts with Cartesian decomposition \citep{Gilyen2019} of $\mathbf{M}_k$:
\begin{align}
    \mathbf{M}_k=L+iH,
\end{align} where the real and imaginary parts of the Hermitian matrices are
\begin{align}\label{eq:LM}
    L=\frac{\mathbf{M}_k+\mathbf{M}_k^\dagger}{2}, H=\frac{\mathbf{M}_{k}-\mathbf{M}_k^\dagger}{2i}.
\end{align}
assuming that $L$ is negative definite  guarantees the asymptotic stability of the dynamics. Under this stability assumption, the non-unitary evolution operator of \cref{eqn:cre_solu} can be expressed as linear combination of Hamiltonian simulation problems:      
    \begin{equation}\label{eqn:LCHS_intro}
      \mathcal{T} e^{-\int_0^t \mathbf{M}_k \ud s} = \int_{\mathbb{R}} \frac{f(k)}{ 1-ik} \mathcal{T} e^{-i \int_0^t (kL+H) \ud s} \ud k.
    \end{equation}
    where $f(k)$ is a kernel function with the form:
\begin{equation}\label{eqn:kernel_intro}
    f(z) = \frac{1}{2\pi e^{-2^\beta} e^{(1+iz)^{\beta}} }, \quad \beta \in (0,1)
\end{equation}
that is near-optimal for LCHS. 
The discretized approximation of \cref{eq:Zk} at the final time $t$ is,
\begin{equation}\label{eqn:CRD_solu_discrete}
  \mathbf{Z}_k(t) \approx \sum_{j=0}^{M-1} c_j U(t, 0, k_j) \ket{\mathbf{Z}(0)} +  \sum_{j'=0}^{M'-1} \sum_{j=0}^{M-1} c'_{j} c_{j} U(t,s_{j'},k_{j})  \ket{b(s_{j'})}. 
\end{equation}
where 
\begin{equation}\label{eq:UTSK}
U(t,s,k) = \mathcal{T} e^{-i \int_s^t (kL(s')+H(s')) \ud s'}. 
\end{equation}
and $M$ is the total number of the nodes.
Given the circuit for the unitaries $U(t,s,k)$, each summation can be implemented by the linear combination of unitaries (LCU) technique~\cite{childs_hamiltonian_nodate} for both terms in \cref{eqn:CRD_solu_discrete} and the sum of the two terms can be computed by another LCU. 
Since $\mathbf{M}_k$ is time-independent, we use quantum singular value transform (QSVT) \citep{gilyen_quantum_2019} as the Hamiltonian simulation subroutine to implement each $U(T,s,k)$ with the lowest asymptotic complexity \citep{an_linear_2023}. 
We start with constructing the select oracle of $U(t, 0, k_j)$ (the key to implementing the LCU) by directly block-encoding the time-independent $\mathbf M_k$ as $\text{BE}_{M_{k}}$,
\begin{equation}\label{eq:alpha-Mk}
  \left( \bra{0}_a \otimes I \right) \text{BE}_{\mathbf M_k}  \left( \ket{0}_a \otimes I \right) = \frac{\mathbf M_k}{\alpha_{\mathbf M_k}}. 
\end{equation}
where $\bra{0}_{a}$ 
is the ancilla register for the block-encoding of $\mathbf M_k$, $\text{BE}_{\mathbf M_k}$ and $\alpha_{\mathbf M_k} \ge \|\mathbf M_k\| $ is the block-encoding factor of $\mathbf M_k$.
We then construct the block-encoding of $kL+H$ in \cref{eq:UTSK} such that 
\begin{equation}
    (\bra{0}_{a'}\otimes I) \text{HAM-T}_{kL+H} (\ket{0}_{a'} \otimes I) = \sum_{j=0}^{M-1}\ket{j}\bra{j} \otimes \frac{k_j L + H}{\alpha_L K + \alpha_H }. 
\end{equation}
Next, for each realization of LCU, we use QSVT~\cite[Corollary 60]{gilyen_quantum_2019} to obtain the select oracle of $U(t, 0, k_j)$, i.e., queries to the block-encoding of $\mathbf M_k$, 
\begin{equation}\label{eq:sel_oracle}
    \text{SEL} = 
    \mathcal{O}\left( (\alpha_LK+\alpha_H) t + \log(1/\epsilon_1)\right) = \mathcal{O} \left( \alpha_{\mathbf M_k} K t + \log(1/\epsilon_1) \right),
\end{equation}
where $\epsilon_1$ represents $\epsilon_1$-precise Hamiltonian simulation. 
To prepare the state of $\int_0^t e^{-\mathbf{M}_k(t-s)} \mathcal B \ud s$ due to the time-independent inhomogeneous term $\mathcal B$, we use QSVT for Hamiltonian simulation to construct
the selection operator for each LCU run:
\begin{equation}\label{eqn:algorithm_inhomo_sel_HS}
    \text{SEL}' = \sum_{j'=0}^{M'-1} \sum_{j=0}^{M-1} \ket{j'} \bra{j'} \otimes \ket{j} \bra{j} \otimes W_{j,j'}, 
\end{equation}
where $W_{j,j'}$ is the block-encoding of $e^{-i (t-s_{j'}) (k_j L + H) }$. 
The error arises from the block-encoding procedure is \citep[Definition 43]{gilyen_quantum_2019} 
\begin{equation}\label{eq:alpha-Mk-error}
 \| \alpha_{\mathbf{M}_k}\left[ \mathbf M_k - \left( \bra{0}_a \otimes I \right) \text{BE}_{\mathbf M_k}  \left( \ket{0}_a \otimes I \right) \right] \| \le \epsilon_{\text{BE}}. 
\end{equation}
Then, we can provide a robust block-Hamiltonian simulation by \citet[Corollary 62]{gilyen_quantum_2019}:
\begin{lemma}\label{lemma:blockHamSimRob}
    Let $t\in\R$, $\eps_{\text BE}\in(0,1)$ and let $W$ be an $(\alpha,a,\eps/|2t|)$-block-encoding of the Hamiltonian $U$. Then we can implement an $\eps_{\text BC}$-precise Hamiltonian simulation unitary $V$ which is an $(1,a+2,\eps_{\text BC})$-block-encoding of $e^{i t W}$, with $6\alpha|t|+9\log(12/\eps)$
	uses of $W$ or its inverse, $3$ uses of controlled-$W$ or its inverse, using $\bigO{a\left(\alpha|t|+\log(2/\eps)\right)}$ two-qubit gates and using $\bigO{1}$ ancilla qubits.
\end{lemma}
\begin{proof}
\cref{lemma:blockHamSimRob} is a direct result of \citet[Lemma 61]{gilyen_quantum_2019}. 
\end{proof}	
    
We now state the complexity of solving \cref{eq:Zk} following time-independent LCHS of \citet[Corollary 16]{an_linear_2023}.
\begin{lemma}[Complexity of the RDE solution using LCHS, adapted from Theorem 14 in \citet{an_quantum_2023}]\label{cor:complexity_homo_time_independent}
  Consider the ODE system in~\cref{eq:Zk} noting time-independent $\mathbf{M}_k$ and $\mathcal{B}$.  
  Suppose that $L$ in \cref{eq:LM} is positive semi-definite, and we are given the block-encoding oracle for the time-independent coefficient matrix $\mathbf M_k$, state preparation oracles for the initial condition $\mathbf Z_k(0)$ and time independent inhomogeneous term $\mathcal B$ (the explicit form of all oracles are given below). 
    Let $\alpha_M \geq \|\mathbf{M}_k\|$ be the block-encoding factor of $\mathbf{M}_k$, we can prepare an $\epsilon$-approximation of the normalized solution $\ket{\mathbf{Z}_k(t)}$ with $\Omega(1)$ probability and a flag indicating success, using
        \begin{equation}
	  \widetilde{\mathcal{O}}\left( \frac{\|\mathbf{Z}_k(0)\| + t\| \mathcal B \|}{\|\mathbf{Z}_k(t)\|} \alpha_{\mathbf M_k} t \left(\log\left(\frac{1}{\epsilon}\right)\right)^{1/\beta} \right)
        \end{equation}
        queries to the block-encoding of $\mathbf{M}_k$, and
        \begin{equation}
            \mathcal{O}\left( \frac{\|\mathbf{Z}_k(0)\| + t\| \mathcal B \|}{\|\mathbf{Z}_k(t)\|}  \right)
        \end{equation}
        queries to the state preparation oracles
	\begin{align}
	  O_{\mathbf Z}: & \ket{0} \rightarrow \ket{\mathbf Z_k(0)}  \\
	  O_{\mathcal B}:&  \ket{0} \rightarrow \ket{\mathcal B}.
	\end{align}
$\beta \in(0, 1)$ defined in \cref{eqn:kernel_intro} determines the asymptotic convergence performance of the kernel function for LCHS.    
\end{lemma}
\begin{proof}
  For each realization of LCU, we use $1$ query to $O_{\mathbf Z}$, and \cref{eq:sel_oracle} queries to the block-encoding of $\mathbf M_k$. Defining the dissipation parameter
\begin{equation}
   g =  \frac{\|\mathbf{Z}_k(0)\| + t\| \mathcal B \|}{\|\mathbf{Z}_k(t)\|},
\end{equation}
the error tolerance $\epsilon_1$ of Hamiltonian simulation in \cref{eq:sel_oracle} can be chosen as 
\begin{equation}\label{eq:eps1}
\epsilon_1 = \frac{1}{8\|c\|_1 g}\epsilon
\end{equation}
and the truncation threshold as 
\begin{equation}\label{eq:K-threshold}
K = \mathcal{O}\left( \left(\log\left(\frac{g}{\epsilon}\right)\right)^{1/\beta} \right) 
\end{equation}
according to \citet[Eq.123\&127]{an_linear_2023}. 
Plugging \cref{eq:eps1} and \cref{eq:K-threshold} into \cref{eq:sel_oracle}, we obtain  
the query complexity to the block-encoding of $\mathbf M_k$ for each run of LCU: 
    \begin{equation}
      \mathcal{O} \left( \alpha_{\mathbf M_k} t  \left(\log\left(\frac{g}{\epsilon}\right)\right)^{1/\beta} + \log\left(\frac{\|c\|_1 g}{\epsilon}\right) \right) = \mathcal{O} \left( \alpha_{\mathbf M_k} t  \left(\log\left(\frac{g}{\epsilon}\right)\right)^{1/\beta} \right). 
    \end{equation}
    The number of repetitions until success is $\mathcal{O}(g)$ according to \citet[Eq.131]{an_linear_2023}, which leads to
    \begin{align}
      \text{query complexity} & = \mathcal{O} \left( g \alpha_{\mathbf M_k} t  \left(\log\left(\frac{g}{\epsilon}\right)\right)^{1/\beta} \right) \\ 
      \label{eq:qc-LCHS}
                              & = \mathcal{\tilde O} \left( g \alpha_{\mathbf M_k} t  \left(\log\left(\frac{1}{\epsilon}\right)\right)^{1/\beta} \right). 
    \end{align}
\end{proof}

\begin{corollary}
  Under the assumption of \cref{cor:complexity_homo_time_independent}, the query complexity can be further bounded by the explicit form of the block-ecoding factor in terms of the Carleman matrix norm $\alpha_{\mathbf M_k} = C_{\rm BE} \|\mathbf M_k\|$,
\begin{equation}\label{eq:qc-LCHS-RD}
\text{query complexity} =
  \widetilde{\mathcal{O}}\left( g k \left[ 4 d n^2 \max_{i\in1, \dots, S} D_i + \max_{i\in1, \dots, S}|\mu_i| + \sum_{i=2}^\varsigma \mathcal C_{\max, i} s_i \right] t \left(\log\left(\frac{ 1 }{ \mathcal{E}}\right)\right)^{1/\beta}  \right).
\end{equation}
\end{corollary}

\begin{proof}
First we consider the $\alpha_M$ term, which can be lower bounded by the spectral norm of the $k^{th}$-order Carleman matrix:
\begin{equation}\label{eq:M-norm-form}
\|\mathbf{M}_k\| = \biggl\|\sum_{j=1}^{k}|j\rangle\langle j|\otimes \mathbf{B}_{j}^j+\sum_{j=1}^{k-i+1}|j\rangle\langle j+i-1|\otimes \mathbf{B}_{j+i-1}^j\biggr\| \le k(\|\mathbf{\tilde{F}}_1\|+ \sum_{i=2}^\varsigma \|\tilde{\mathbf{F}}_i\|).
\end{equation}
The norm of the high-order ($\varsigma \ge 2$) coefficient matrix (\cref{eq:spatial_tensor_formula}) are given by \cref{eq:tildeF-norm}.
The final bound of $\norm{\mathbf{M}_k}$ for a $S$-species and $\varsigma$-order reaction diffusion system can be obtained by plugging \cref{eq:tildeF1-norm} and \cref{eq:tildeF-norm}  into \cref{eq:M-norm-form},
\begin{equation}\label{eq:M-norm}
  \|\mathbf{M}_k\| \le k \left[ 4 d n^2 \max_{i\in1, \dots, S} D_i + \max_{i\in1, \dots, S}|\mu_i| + \sum_{i=2}^\varsigma \mathcal C_{\max, i} s_i \right].
\end{equation}
For an autocatalytic reaction system (e.g., the GM model), \cref{eq:M-norm} can be further simplified to
\begin{equation}\label{eq:M-norm-auto}
  \|\mathbf{M}_k\| \le k \left[ 4 d n^2 \max_{i\in1, \dots, S} D_i + \max_{i\in1, \dots, S}|\mu_i| +  \sqrt{2(2S-1)}\sum_{i=2}^\varsigma \mathcal C_{\max, i} \right].
\end{equation}
Finally, the query complexity to the block-encoding of $\mathbf M_k$ can be obtained by plugging \cref{eq:M-norm-auto} into \cref{eq:qc-LCHS}, which yields \cref{eq:qc-LCHS-RD} and complete the proof.
The constant $C_{\rm BE} > 1$ is absorbed in $\widetilde{\mathcal{O}}$.
\end{proof}
The quantum query complexity (\cref{eq:qc-LCHS-RD}) offers a quadratic scaling in 1-D number of grid points $n$ and polynominal scaling in number of species assuming the coefficient matrices in \cref{eq:discrete_Turing} is $s=\polylog (S)$-sparse. This indicates a potential exponential speedup in complexity compared to classical algorithms that scales exponentially with number of degrees of freedom, i.e., $O(S^{\varsigma n_d})$, in solving \cref{eq:discrete_Turing} directly. However, it is possible that classical Monte-Carlo simulations or future optimization of classical algorithms could offer a better-than-exponential cost.      

With quantum algorithms for both the reaction rate calculation and the
nonlinear reaction-diffusion equation in hand, we are now ready to present the
final full quantum complexity for simulating the reaction-diffusion process.
\begin{theorem}[Final query complexity for combined reaction rate and RDE]
  Under the assumptions of \cref{thm:UC}, \cref{cor:alpha-F-varsigma}, and \cref{cor:complexity_homo_time_independent}, the total queries for blocking-encode the reaction-rate matrix $\tilde{\mathbf F}_j$ and the Carleman matrix $\mathbf M_k$ scales at most
  \begin{equation}\label{eq:final-query-complexity}
    \widetilde{O}\left(  \frac{\max_j\alpha(j)(1+\max_i\alpha(i)/k_BT)\log(\alpha_{\Delta G}/k_BT)\log^2(\sum_{ij}|\alpha(ij)-\beta(ij)|/\epsilon)}{\gamma\delta} \cdot \left[ g \alpha_{\mathbf M_k} t  \left(\log\left(\frac{1}{\epsilon}\right)\right)^{1/\beta} \right] \right)
  \end{equation}
\end{theorem}
\begin{proof}
  The proof is a direct result of \cref{thm:UC}, ~\cref{cor:alpha-F-varsigma}, and~\cref{cor:complexity_homo_time_independent}. The number of queries for blocking encoding $\tilde{\mathbf{F}}_j$ and $\mathbf M_k$ are given by \cref{eq:alpha-reactionRate} and \cref{eq:qc-LCHS}, respectively. Then, the total number of queries is a product of \cref{eq:alpha-reactionRate} and \cref{eq:qc-LCHS}, yielding the final query complexity \cref{eq:final-query-complexity} and completing the proof. 
\end{proof}
The total block encoding can be implemented within error $\alpha_{\mathbf M_k}\epsilon_{\rm BE} + \alpha_F \epsilon$ since the error of the product of block-encoded matrices is simply an add-up of the block encoding error of each matrix  \citep[Lemma 53]{Gilyen2019}. Here, the block encoding factors $\alpha_F$ and $\alpha_{\mathbf M_k}$ are given in \cref{eq:alpha-F-varsigma} and \cref{eq:alpha-Mk}, respectively. Their block encoding error are given in \cref{eq:alpha-F-error} and \cref{eq:alpha-Mk-error}.
The results show that the query complexity of the overall simulation for computing the solution to the RDE scales logarithmically with the block encoding factors of calcualting the reaction rate and polynomially with the number of grid points $n_d$ and species $S$ for the nonlinear RDE under the assumptions of \cref{thm:UC}, \cref{cor:alpha-F-varsigma}, and \cref{cor:complexity_homo_time_independent}. This suggests an exponential speedup compared with classical algorithms.  

\subsection{Numerical Simulations} \label{sec: simulations}
In this section, we examine the convergence behavior of the Carleman linearization method through numerical simulations across different stability regimes of the GM model, as described in \cref{sec:GM}. Specifically, we solve the Carleman-linearized form of the GM model (\cref{eq:Zk}; hereafter referred to as ``Carleman-GM'') for various truncation orders, $k$, and compare these solutions with the direct numerical simulation of \cref{eq:GM_a_rho1} and \cref{eq:GM_s_rho1}, referred to as ``RK4-GM''. The simulations were performed using physically relevant parameter scales, including time steps, spatial discretization, diffusion coefficients, and other model parameters, as detailed in \cref{tab:scales}. Both the GM model and the Carleman-GM system are solved using the Runge-Kutta 4th-order (RK4) time-stepping scheme to ensure consistency in numerical methods. The simulations use a time step satisfying Courant-Friedrichs-Lewy conditions for numerical convergence.
The Carleman-GM solution begins to converge toward the RK4-GM solution at truncation order $k=3$ as demonstrated by \cref{fig:domain}, which shows the absolute error ${\norm{\mathcal{E}_{abs}}}_\infty$ of the Carleman-GM  at different orders compared to the RK4-GM solution.  
\newcommand\fx{10}
\newcommand\fy{42.5}
\begin{figure}
    \centering
\begin{overpic}[width=\textwidth]{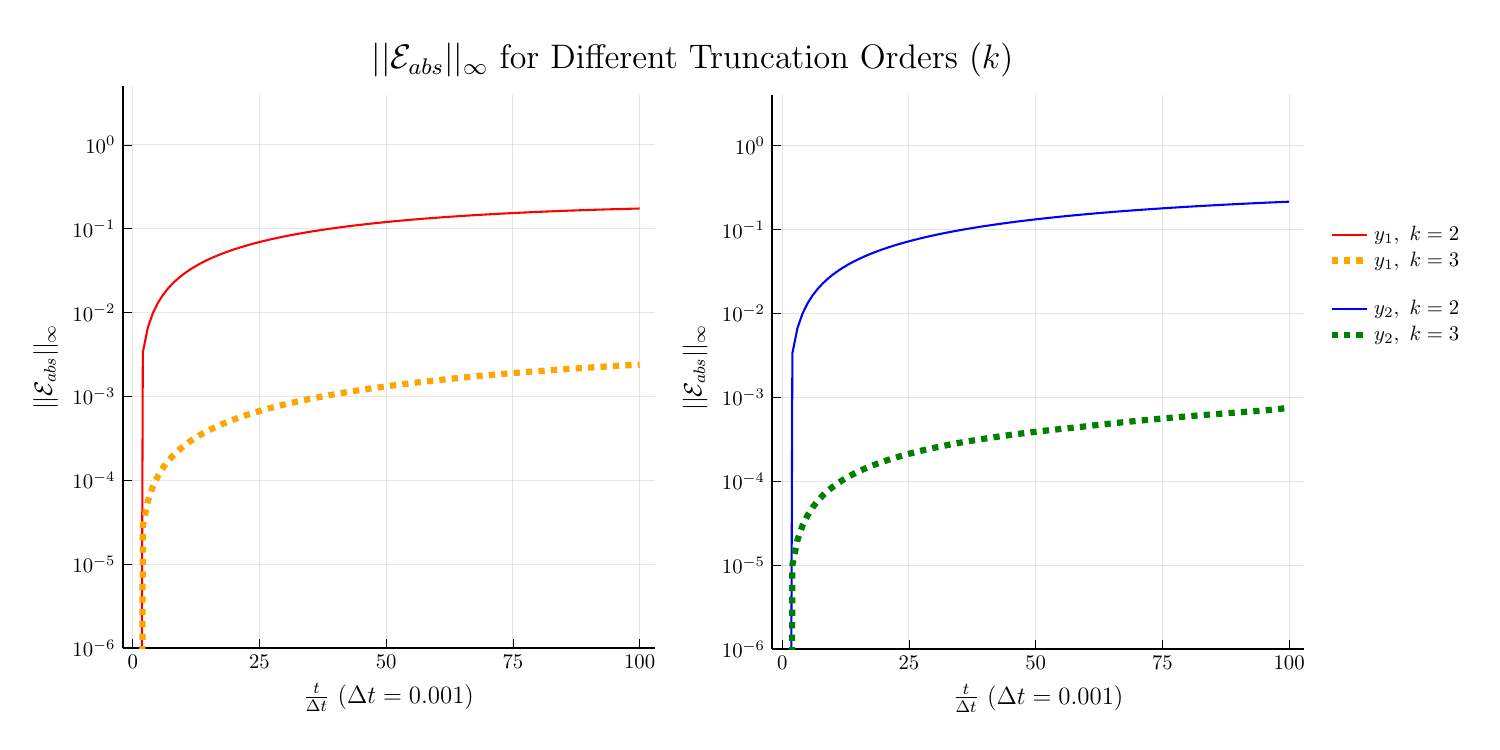}\put(\fx,\fy){(a)}\put(53,\fy){(b)}\end{overpic}
    \caption{Absolute error ${\norm{\mathcal{E}_{abs}}}_\infty$ between RK4-GM and the Carleman-GM with Carleman truncation order $k=2$ and $k=3$ for (a): $y_1$ and (b): $y_2$. Values of the dimensionless parameters in \cref{eq:GM_a_rho1} were chosen to represent a typical behavior for the GM system in a stable regime with $D_1=1\times 10^{-4}, D_2=D_1/2, \mu_1=\mu_2=5, c_1=b_1=1,$ and $b_2=0$. The initial conditions are $\mathbf{Y}_0=[1 + \sin(2\pi x_i), 1 + \cos(4\pi x_i)]$ for $i = 1, 2, \ldots, 50$ spatial nodes. The time step is set to $dt=0.001$.}
    \label{fig:domain}
\end{figure}
\newcommand\fpx{2}
\newcommand\fpy{94}
\begin{figure}
    \centering 
    \begin{overpic}[width=0.8\textwidth]{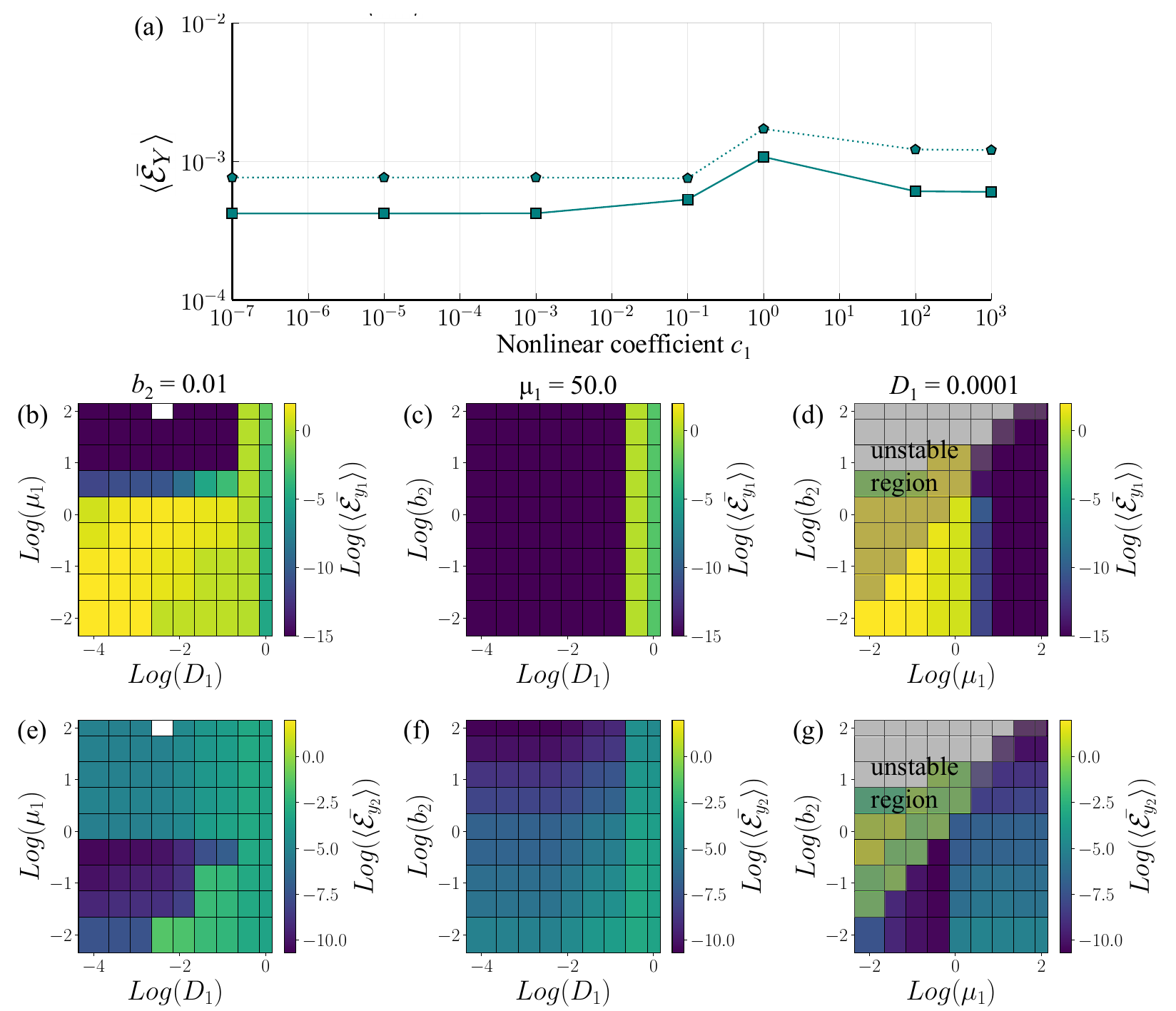}
    \end{overpic}
    \caption{Relative Carleman truncation error $\langle \bar{\mathcal{E}}_Y \rangle$ for different parameter regimes. Panel (a) varies the nonlinear coefficient $c_1$ in \cref{eq:GM_a} (solid line) and \cref{eq:GM_s} (dotted line) with $D_1=3\times10^{-4}\, \text{and}\, D_2=2\times 10^{-5}$, and the other parameters are set to 1. For panel (b)-(g), we use \cref{scaled} and \cref{scaled1} with characterized stability. 
    We plot the slice with inhomogeneous parameter $b_2 = 0.01$ for panel (b) and (e), linear decaying rate $\mu_1 = 50.0$ for (c) and (f), and diffusion coefficient $D_1 = 10^{-4}$ ($D_2=D_1/2$) for (d) and (g). Panel (b)-(d) is the solution corresponding to $y_1$ while (e)-(g) is the solution for $y_2$. 
    }
    \label{fig:combined}
\end{figure}

A crucial question is how Carleman truncation error behaves with the strength of nonlinearity. To address this, we examine the relative error $\mathcal{{E}_\mathbf{Y}}$ as a function of the nonlinear coefficient $c_1$ in \cref{eq:GM_a_rho1} and \cref{eq:GM_s_rho1}. Here, the relative error is defined as $\mathcal{{E}_\mathbf{Y}}= (\mathbf{Z_k}-\mathbf{Y}_{RK4}) / \mathbf{Y}_{RK4}$ with $\mathbf{Y}_{RK4}=[y_{1, RK4}, y_{2, RK4}
]^\top$ being the RK4-GM solutions. The domain and time averaged relative error $\langle \bar{\mathcal{E}}_Y \rangle$ remains around $10^{-3}$ across the parameter space of $c_1$ (\cref{fig:combined}(a)), suggesting the applicability of Carleman linearization for RDE in the nonlinear regime explored here.
We further explore how the Carleman truncation error behaves in stable and unstable regimes of RDEs characterized by interactions among the diffusion, linear decay, nonlinear reaction, and inhomogeneous terms. Specifically, we examine how $\langle \bar{\mathcal{E}}_Y \rangle$ responds to biologically relevant parameter ranges of $D_1$, $\mu_1$, and $b_2$ in the following  
rescaled GM model that reduces the number of parameters \citep{Wu2017}: 
\begin{align}\label{scaled}
    \frac{\partial y_1}{\partial t} &= D_1\nabla y_1 - \mu_1 y_1 + y_1^2 y_2, \\
    \frac{\partial y_2}{\partial t} &= D_2 \nabla y_2 - y_2 - y_1^2 y_2  + b_2. \label{scaled1}
\end{align}
Two equilibrium points arise from \cref{scaled1} when $0<\mu_1/b_2<2$ \citep{Wu2017}, which divides the RK4-GM system into stable and unstable regions (marked by shade in \cref{fig:combined}(d) and (g)). The Carleman truncation error for the inhibitor, $\langle \bar{\mathcal E}_{y_2}\rangle$, is on the order of $10^{-3}$ or smaller in all parameter ranges (\cref{fig:combined}(e)-(f) and unshaded area in \cref{fig:combined}(g)) other than in the unstable region (shade in \cref{fig:combined}(g)). This is likely because the inhibitor $y_2$ in \cref{scaled1} is \textit{not} governed by a self nonlinear equation. The Carleman truncation error for the activator, $\langle \bar{\mathcal E}_{y_1}\rangle$, is far more complex. For a given inhomogeneous source term $b_2 = 0.01$ and fixed coefficient $c_1 = 1$ of the nonlinear reaction term, $\langle \bar{\mathcal E}_{y_1}\rangle$ decreases with increasing diffusion coefficient $D_1$ and linear decay rate $\mu_1$ (\cref{fig:combined}(b)). Larger decay rate $\mu_1$ suppresses $\langle \bar{\mathcal E}_{y_1}\rangle$ machine error due to the suppression of nonlinear reaction terms as further evidenced in \cref{fig:combined}(c) and the stable region of \cref{fig:combined}(d) when $\log \mu_1 > 0$. However, $\langle \bar{\mathcal E}_{y_1}\rangle$ becomes prominent when both $\mu_1$ and $D_1$ are small (\cref{fig:combined}(b)), even in the stable region (shade region in \cref{fig:combined}(d)). Overall, the Carleman truncation for the activator is determined by complex interactions among the linear decay rate, diffusion coefficient, and the inhomogeneous source term while the one for the inhibitor is largely determined by the stability of the RDE system. However, the Carleman truncation error for both activator and inhibitor is acceptable for biologically relevant parameter regime featured by weak nonlinearity. In fact, it is the vast number of degrees of freedom that hinders simulating the biosystems at their native scales rather than the nonlinearity. This makes quantum algorithms an idea candidate to tackle the dimensionality curse of biosystems.

\section{Conclusion}

Simulating realistic microbial systems across a wide range of spatiotemporal
scales is crucial, yet, classically intractable. This is the so-called curse of
dimensionality. In this work, we aim to break this curse using quantum
computing as it promises an exponential speedup in solving certain classes of
problems. Specifically, we explore potential quantum advantage in solving
reaction-diffusion equations (RDEs) that govern the dynamics of a wide class of
biosystems. Given a $S$-species and $\varsigma$-order RDE system, the complexity from classical algorithms scales exponentially with number of degrees of freedom, i.e., $O(S^{\varsigma n_d})$ with $n_d$ being the number of grid points. This makes simulating the RDE system prohibitively expensive on classical computers. More importantly, determining the reaction rate at biological scales is classically NP hard.
We overcome these challenges by harnessing the power of quantum computing. First, we
formulated a generalized framework of RDE system with arbitrary number of reacting species ($S$) and order of nonlinear reaction ($\varsigma$) to exploit the exponential quantum advantage in Hilbert space. Then, we calculated the reaction rate related to free energy differences in polynomial time, which is an exponential speedup compared to classical algorithms as reaction rates are not believed to be computable in polynominal time on a classical computers. 

Finally, we solved the $S$-species and $\varsigma$-order nonlinear RDEs in polynomial time. This is done by first Carleman-linearizing the $S$-species and $\varsigma$-order RDE system and then apply linear combination of Hamiltonian simulations (LCHS) to the linearized system. We found that the Carleman truncation errors converges for the majority of biologically relevant parameters in the stable regime by performing numerical simulations of the canonical Grierer-Meinhardt (GM) model. We then recast the linearized system into LCHS, each Hamiltonian of which was solved by quantum singular value transform. The quantum query complexity to the block-encoding of the Carleman matrix scales polynominally with $S$ and $n_d$. 
The final query complexity for combined reaction rate and reaction-diffusion equation therefore scales logarithmically with block-encoding factors for reaction rate and polynomially with $S$ and $n_d$.

Future work will focus on providing a tight Carleman truncation error bounds related to the stability of $S$-species and $\varsigma$-order RDE systems. In addition, efficient state preparation and measurements are urgently needed to obtain an end-to-end quantum advantage even though they are bottleneck problems in quantum computing. Furthermore, robust cost estimation including quantum error correction and energy consumption will be crucial for exploring the potential quantum advantage for multiscale biosystems.      

The application of this framework on suitable quantum computing platforms has the potential to extend current cell modeling capabilities, leading to improved understanding of fundamental biological processes and emergent phenomena such as drug delivery, disease dynamics, and resource allocation. By enabling faster predictions and enhancing scalability, quantum computing could accommodate simulations of greater complexity and larger scale, surpassing the computational limitations of classical approaches. The methodology developed in this study is broadly applicable to other nonlinear systems, including other reaction-diffusion models, such as the Lorenz attractor, Lotka-Volterra equations, and reactive Langevin systems. Additionally, this framework could be adapted to stochastic chemical systems derived from Chemical Master Equations, which involve high-dimensional spaces with interactions across numerous species and reactions. As quantum computing technology continues to advance, its integration into biological research may provide solutions to computational challenges in areas such as whole-cell modeling, drug discovery, and microbial ecology. By addressing these previously intractable problems, quantum computing has the potential to support a deeper understanding of complex and multiscale biological systems.

\section{Acknowledgments}

This research was supported by Pacific Northwest National Laboratory's Quantum Algorithms and Architecture for Domain Science (QuAADS) Laboratory Directed Research and Development (LDRD) Initiative. XYL and NW also acknowledge the support from DOE, Office of Science, National Quantum Information Science Research Centers, Co-design Center for Quantum Advantage (C2QA) under Contract No.~DE-SC0012704 (Basic Energy Sciences, PNNL FWP 76274). This work is also partially supported by the NW-BRaVE for Biopreparedness project funded by the U. S. Department of Energy (DOE), Office of Science, Office of Biological and Environmental Research, under FWP 81832. PNNL is a multi-program national laboratory operated by Battelle for the DOE under Contract DE-AC05-76RL01830. This work was supported in part by the U.S. Department of Energy, Office of Science, Office of Workforce Development for Teachers  Scientists (WDTS) under the Science Undergraduate Laboratory Internships Program (SULI). We thank Xiaolong Yin for discussion of the potential application of Lattice Boltzmann Method for RDEs and Dong An for discussing LCHS.

\clearpage
\clearpage

\appendix

\section{Polynomial structure of the reaction rate}
\label{sec:poly-structure}

For example, a system has mixed orders:${\varsigma_1, \varsigma_2, \varsigma_3} = {1, 2, 3}$ when  
\begin{align}
\text{Reaction 1:} \quad &A \rightarrow B \quad (\varsigma_1 = 1) \\
\text{Reaction 2:} \quad &A + B \rightarrow C \quad (\varsigma_2 = 2) \\
\text{Reaction 3:} \quad &2A + B \rightarrow D \quad (\varsigma_3 = 3) \quad .
\end{align}
\begin{enumerate}
  \item Binary collision between one A molecule and one B molecule 
\begin{align}
\text{Reaction:} \quad &A + B \rightarrow C \\
\text{Stoichiometry:} \quad &\alpha_A = 1, \alpha_B = 1 \\
\text{Order:} \quad &\varsigma_r = 1 + 1 = 2 \\
\text{Rate law:} \quad &\text{Rate} = c_{A, B}[A][B] \quad \text{(2nd degree polynomial)}
\end{align}

\item Termolecular collision involving two A molecules and one B molecule 
\begin{align}
\text{Reaction:} \quad &2A + B \rightarrow \text{products} \\
\text{Stoichiometry:} \quad &\alpha_A = 2, \alpha_B = 1 \\
\text{Order:} \quad &\varsigma_r = 2 + 1 = 3 \\
\text{Rate law:} \quad &\text{Rate} = c_{2A, B}[A]^2[B] \quad \text{(3rd degree polynomial)}
\end{align}

\item Complex stoichiometry ($\varsigma_r = 6$)
\begin{align}
\text{Reaction:} \quad &3A + 2B + C \rightarrow \text{products} \\
\text{Stoichiometry:} \quad &\alpha_A = 3, \alpha_B = 2, \alpha_C = 1 \\
\text{Order:} \quad &\varsigma_r = 3 + 2 + 1 = 6 \\
\text{Rate law:} \quad &\text{Rate} = c_{3A, 2B, C}[A]^3[B]^2[C] \quad \text{(6th degree polynomial)}
\end{align}
\end{enumerate}

\section{Autocatalytic reaction}\label{sec:autocatalytic}
We take the autocatalytic reaction as an example because biological reproduction rests ultimately on chemical autocatalysis \citep{konnyu_kinetics_2024}. A reaction is autocatalytic in species $i$ if:
\begin{enumerate}[label=(\roman*)]
    \item Species $i$ appears as both reactant and product: $\alpha_{r,i} > 0$ and $\beta_{r,i} > 0$
    \item Net production of species $i$: $\beta_{r,i} > \alpha_{r,i}$
    \item Catalytic dependence: The rate increases with $[y_i]$
\end{enumerate}
The canonical autocatalytic reaction of order $\Order$ is:
\begin{align}
(\Order-1)y_i + y_j \xrightarrow{c_{ij}} \Order y_i
\end{align}
with stoichiometry $\alpha_{r,i} = \Order-1, \alpha_{r,j} = 1, \beta_{r,i} = \Order, \beta_{r,j} = 0$.
For pure autocatalytic networks where each reaction has the form $(\Order-1)y_i + y_j \rightarrow \Order y_i$, the coefficient tensor simplifies to:
\begin{align}\label{eq:autocatalytic_F_varsigma}
\mathbf{F}_\Order[m, \ell] = \begin{cases}
c_{ij} & \text{if } m = i \text{ and } \ell = \ell_{\text{canon}}(i,j) \\[0.8em]
-c_{ij} & \text{if } m = j, \, i \neq j \text{ and } \ell = \ell_{\text{canon}}(i,j) \\[0.8em]
0 & \text{otherwise}
\end{cases}
\end{align}
where $\ell_{\text{canon}}(i,j) = \IndexFunc_\Order(\underbrace{i,i,\ldots,i}_{\Order-1}, j)$ is the canonical position, the explicit form of which is given by
\begin{align}\label{eq:canonical_position_formula}
\ell_{\text{canon}}(i,j) = j + (i-1) \cdot \frac{S(S^{\Order-1} - 1)}{S - 1}
\end{align}
for $S \geq 2$ and $\Order \geq 1$.
\cref{eq:autocatalytic_F_varsigma} is a direct result of \cref{eq:general_F_varsigma} considering
\begin{enumerate}
\item net change in species $i$: $\beta_{r,i} - \alpha_{r,i} = \Order - (\Order-1) = 1$
\item net change in species $j$ (if $j \neq i$): $\beta_{r,j} - \alpha_{r,j} = 0 - 1 = -1$  
\item net change in species $j$ (if $j = i$): $\beta_{r,j} - \alpha_{r,j} = \Order - \Order = 0$
\end{enumerate}
For a two-species ($S=2$) reaction-diffusion system,
we have 
\begin{equation}\label{eq:Y^3}
    \mathbf{Y}^{\otimes 3} = [y_1^3, y_1^2 y_2, y_1 y_2 y_1, y_1 y_2^2, y_2 y_1^2, y_2 y_1 y_2, y_2^2 y_1, y_2^3]^\top.
  \end{equation}
The tensor positions for different reactions are:
\begin{align}
\begin{array}{c|c|c}
\text{Reaction} & \text{Canonical Ordering} & \text{Position} \\
\hline
2y_1 + y_1 \rightarrow 3y_1 & y_1^3 & \{1\} \\
2y_1 + y_2 \rightarrow 3y_1 & y_1^2 y_2 & \{2\} \\
2y_2 + y_1 \rightarrow 3y_2 & y_2^2 y_1 & \{7\} \\
2y_2 + y_2 \rightarrow 3y_2 & y_2^3 & \{8\}
\end{array}
\end{align}
The resulting $F_3 \in \mathbb{R}^{2 \times 8}$ is:
\begin{align}
F_3 = \begin{bmatrix}
c_{11} & c_{12} & 0 & 0 & 0 & 0 & c_{21} & 0 \\
0 & -c_{12} & 0 & 0 & 0 & 0 & -c_{21} & c_{22}
\end{bmatrix}
\end{align}
For pure autocatalytic networks with maximum rate constant $c_{\rm max} = \max_{i,j} c_{ij}$:
\begin{align}\label{eq:F-sigma-l2}
\|\mathbf{F}_\Order\|_2 \leq c_{\rm max}\sqrt{2(2S-1)}
\end{align}
because $\sigma_{\max} = 2S-1$ (maximum row sum) and $\tau_{\max} = 2$ (maximum
column sum).

\section{Carleman Truncation with Example}\label{sec:example}
In this section we include more practical details on how to implement the Carleman method with an example for the GM model. Recall the RDEs model with the linearized dissipative term, \cref{eq:discrete_Turing}. For the GM model, we can compactly denote this nonlinear spatial-discrete system with inhomogenous terms:
\begin{align}
\frac{dy_1}{dt} = \alpha + \beta y_1 + \gamma y_2+ \xi y_1^2 y_2 \label{eq:form1}\\
\frac{dy_2}{dt} = \zeta + \eta y_1 + \kappa y_2+ \psi y_1^2 y_2\label{eq:form2}
\end{align}
where
\begin{align}
    \alpha &=  b_1, \ \label{eq:alpha1}\\
    \beta &= D_1I \Delta_h -\mu_1 I, \ \label{eq:beta}\\
    \gamma&= 0I, \ \label{eq:gamma}\\
    \xi &= c_1 I, \ \label{eq:xi}\\
    \zeta &= b_2  \label{eq:zeta}\\
    \eta &= 0I, \ \label{eq:eta}\\
    \kappa &= D_2I\Delta_h  -\mu_2 I, \ \text{and} \ \label{eq:kappa}\\
    \psi &= -c_1 I. \ \label{eq:psi}
\end{align}
Following the spatial discretization, the input parameters $\alpha$, $\zeta$, $\beta$, $\eta$, $\gamma$, $\kappa$, $\xi$, and $\psi$ are all contained within matrices of $k^d$ dimensions. Once constructed, the size of the transfer matrix will be $k^d(2^{n+1}-2)$. All except the initial conditions for $y_1$ and $y_2$ are most likely uniform matrices, though they could be of another form. For instance, the generating or dissipating coefficients could vary in space due to biological factors. 
Consider for the GM model, where the highest polynomial degree is $\varsigma = 3$, we have 
\begin{align}\label{kro_ex}
\mathbf{Y}^{\otimes 1} &= \begin{bmatrix}
  y_1& y_2
\end{bmatrix}^\mathrm{T}, \\ \label{kro_ex2}
\mathbf{Y}^{\otimes 2} &= \begin{bmatrix}
    y_1^2 & y_1y_2 & y_2y_1 & y_2^2
\end{bmatrix}^\mathrm{T}, \text{ and } \\ \label{kro_ex3}
\mathbf{Y}^{\otimes 3} &= \begin{bmatrix}
    y_1^3 & y_1^2y_2 & y_1^2y_2 & y_1y_2^2 & y_1y_2^2 & y_1y_2^2 & y_2^2y_1 & y_2^3
\end{bmatrix}^\mathrm{T}.
\end{align}
\cref{eq:form1} with all terms included can be expressed as 
\begin{align}
\begin{split} \label{eq: example1}
    \frac{dy_1}{dt} &= \alpha + \beta y_1 + \boldsymbol{\gamma}y_2 + 0y_1^2 + 0y_1y_2 + 0y_2y_1 \\
                  &\quad + 0y_1^3 + \xi y_1^2y_2 + 0y_1^2y_2 + 0y_1y_2^2 + 0y_1y_2^2 + 0y_1^2y_2 + 0y_1y_2^2 + 0y_1y_2^2 + 0y_2^3  
\end{split}\\
\begin{split} \label{eq: example2}
    \frac{dy_2}{dt} &= \zeta + \eta y_1 + \kappa y_2+ 0y_1^2 + 0y_1y_2 + 0y_2 y_1 \\
                  &\quad + 0y_1^3 + \psi y_1^2y_2 + 0y_1^2y_2 + 0y_2y_1^2 + 0y_1y_2^2 + 0y_1^2y_2 + 0y_1y_2^2 + 0y_1y_2^2 + 0y_2^3,
\end{split}
\end{align} 
where the nonlinear, quadratic, and cubic terms have zero-valued coefficients. Since the majority of the coefficients in the transfer matrix are zero, the transfer matrix is sparse.
The matrices of coefficients are 
\begin{align} \label{B1}
\mathbf{B}_0^1 &= \begin{bmatrix}\boldsymbol{\beta} & \boldsymbol{\gamma} \\ \boldsymbol{\eta} &\boldsymbol{\kappa}\end{bmatrix}, \\ \label{B2}
\mathbf{B}_1^1 &= \begin{bmatrix}\mathbf{0} & \mathbf{0} & \mathbf{0} & \mathbf{0} \\ \mathbf{0} & \mathbf{0} & \mathbf{0}&\mathbf{0}\end{bmatrix}, \text{ and} \\ \label{B3}
\mathbf{B}_2^1 &= \begin{bmatrix}
\mathbf{0} & \boldsymbol{\xi} & \mathbf{0} & \mathbf{0} & \mathbf{0} & \mathbf{0} & \mathbf{0} & \mathbf{0} \\
\mathbf{0} & \boldsymbol{\psi} & \mathbf{0} & \mathbf{0} & \mathbf{0} & \mathbf{0} & \mathbf{0} & \mathbf{0} \\
\end{bmatrix}, 
\end{align} where $\mathbf{0}$ is a square $k^d \times l^n$ matrix of zeros with $n$ being the number of discrete spatial nodes in each dimension. $\mathbf{B}^1_i$ for $i>2$ where $i \in \mathbb{N}$ is a matrix $2^{i+1} \times 2$ of null matrices $k^d\times n^d$. These matrices are placeholders for higher order polynomials, matching \cref{eq: example1} and \eqref{eq: example2}.
The parameters in \cref{eq:alpha1}-\eqref{eq:psi} are represented inside matrices, $\boldsymbol{\beta}$, $\boldsymbol{\gamma}$, $\boldsymbol{\eta}$, $\boldsymbol{\kappa}$, $\boldsymbol{\xi}$, and $\boldsymbol{\psi}$. These are square $k^d \times n^d$ diagonal matrices with the coefficient value along the diagonal. This is even true for the 1D case, as required to keep the final Carleman matrix a square matrix. Consider if the dimension is $d$ then there are $k^d$ entries, except in the 1D case where there are $2^n$ entries. For the 1D case the input parameters are square diagonal matrices as well, for example let $n=3$, then \begin{align}
    \boldsymbol{\beta}=\begin{bmatrix} \beta&0&0\\0&\beta&0\\0&0&\beta\end{bmatrix}.
\end{align} 
For computational purposes the infinite transfer matrix needs to be truncated to some degree $k$ where $k\in \mathbb{N}$, which introduces an error. 
The system can be truncated to an $k \times n$ matrix of matrices and represented by the differential equations $z=(\mathbf{\tilde{y}}_1,\mathbf{\tilde{y}}_2,...\mathbf{\tilde{y}}_k)$. The truncated system is linear and can therefore be solved with an already developed linear quantum solver. Let the truncated transfer matrix be denoted as 
\begin{align}
\frac{\mathbf{Z}_k^\top}{dt}=\mathbf{M}_k\mathbf{Z}_k^\top +F_0 \label{eq:M2}
\end{align}
where $F_0$ is a vector of constant values with $\boldsymbol{\alpha}$ and $\boldsymbol{\zeta}$ in the first two entries followed by $k^d(2^{n+1}-2)-2l$ 0 entries, such that is has $F_0(t)$ is a column vector with the same length as the transfer matrix. $\boldsymbol{\alpha}$ and $\boldsymbol{\zeta}$ vectors are length $n$ and with each entry value respectively $\alpha$ or $\zeta$. For example for $n=3$, 
\begin{align}
    \boldsymbol{\alpha}=\begin{bmatrix}
        \alpha \\ \alpha \\ \alpha
    \end{bmatrix} \text{ and }
    \boldsymbol{\zeta}=\begin{bmatrix}
    \zeta \\ \zeta \\ \zeta
    \end{bmatrix}.
\end{align}
$\mathbf{M}$ has the block structure:
\begin{align} \label{longM} \mathbf{M}_k=\begin{bmatrix}
\mathbf{B}_0^1 & \mathbf{B}_1^1 & \mathbf{B}_2^1 &\mathbf{0} & \cdots &\mathbf{0} \\
\mathbf{0} & \mathbf{B}_0^2 & \mathbf{B}_1^2 & \cdots &\cdots& \mathbf{B}^2_{n-2} &\\
\mathbf{0} & \mathbf{0} & \mathbf{B}_0^3 & \cdots &\cdots& \mathbf{B}^{3}_{n-3} \\
\vdots & \vdots & \ddots & \ddots & \ddots & \vdots &  \\
\mathbf{0} & \mathbf{0} & \mathbf{0} & \mathbf{0} & \cdots & \mathbf{B}_0^n
\end{bmatrix}.
\end{align}
This is a lower triangular matrix, a square matrix with the value 0 below the diagonal. The matrix is padded with null matrices. 

When we consider that $y_1$ and $y_2$ contain the set of multiple dimensional nodes of the system, the correct implementation of the Kronecker product is important. In the case with multiple discrete spatial nodes, the concentrations have $d$ dimensions and $k^d$ entries. In the 1D case, we have the vectors 
$y_1= \begin{bmatrix}
    y_{1,1} & y_{1,2} & \cdots & y_{1,l}
\end{bmatrix}^{\mathrm{T}}$ and $y_2 = \begin{bmatrix}
    y_{2,1} & y_{2,2} & \cdots & y_{2,l}
\end{bmatrix}^{\mathrm{T}}$.  
The Kronecker products needs to be computed so embedded spatial terms are grouped accordingly, for example:
\begin{align}
  \mathbf{Y}^{\otimes 2} & =
\left[
\begin{array}{cccc}
    y_1^2 & y_1y_2 & y_1y_2 & y_2^2
\end{array}
\right]^{\mathrm{T}} \nonumber \\
			 & =
\left[
\begin{array}{ccccccccccccccccc}
    y_{1,1}^2 & y_{1,2}^2 & \cdots & y_{1,l}^2 & y_{1,1}y_{2,1} & y_{1,2}y_{2,2} & \cdots & y_{1, l}y_{2, 1} &
    y_{2,1}y_{1,1} & y_{2, 2}y_{1,2} & \cdots & y_{2, l}y_{1, l} &y_{2,1}^2 &y_{2,2}^2 & \cdots & y_{2,l}^2
\end{array}
\right]^{\mathrm{T}}.  \nonumber
\end{align}

As an explicit example, consider the GM where the polynomial order is 3 and let $k=3$. The transfer matrix is
\begin{align}\label{eq: transfer_n3}
\mathbf{M}_3=\begin{bmatrix}
        \mathbf{B}_0^1 & \mathbf{B}_1^1 & \mathbf{B}_2^1\\
        \mathbf{0} & \mathbf{B}_0^2 & \mathbf{B}_1^2 \\
        \mathbf{0} & \mathbf{0} & \mathbf{B}_0^3 
    \end{bmatrix}.
\end{align}
The first row of $B$-matrices are the listed in \cref{B1}, \eqref{B2}, and \eqref{B3}. The remaining $B$-matrices are computed:
\begin{align}
\mathbf{B}_0^2=\mathbf{B}_0^1 \otimes \mathbb{I}^{[2-1]}+\mathbb{I}\otimes \mathbf{B}_0^1\\
\mathbf{B}_0^2= \begin{bmatrix}\boldsymbol{\beta} & \boldsymbol{\gamma} \\ \boldsymbol{\eta} &\boldsymbol{\kappa}\end{bmatrix} \otimes \begin{bmatrix}
    1&0\\0&1
\end{bmatrix}+ \begin{bmatrix}
    1&0\\0&1
\end{bmatrix}\otimes  \begin{bmatrix}\boldsymbol{\beta} & \boldsymbol{\gamma} \\ \boldsymbol{\eta} &\boldsymbol{\kappa}\end{bmatrix} \\
\mathbf{B}_0^2=\begin{bmatrix}
        2\boldsymbol{\beta} & \boldsymbol{\gamma} & \boldsymbol{\gamma} & \mathbf{0}\\
        \boldsymbol{\eta} & \boldsymbol{\beta}+\boldsymbol{\kappa} & \mathbf{0} & \boldsymbol{\gamma}\\
        \boldsymbol{\eta} & \mathbf{0} & \boldsymbol{\beta}+\boldsymbol{\kappa} & \boldsymbol{\gamma}\\
        \mathbf{0} & \boldsymbol{\eta} & \boldsymbol{\eta} & 2\boldsymbol{\kappa}\\
    \end{bmatrix},
\end{align}
\begin{align}
\mathbf{B}_1^2=\mathbf{B}_1^1 \otimes I^{[2-1]}+I\otimes \mathbf{B}_1^1\
\end{align}

\begin{align}
    \mathbf{B}_1^2=\begin{bmatrix}\mathbf{0} & \mathbf{0} & \mathbf{0} & \mathbf{0} \\ \mathbf{0} &\mathbf{0}&\mathbf{0}&\mathbf{0}\end{bmatrix} \otimes \begin{bmatrix}
        1&0\\0&1
    \end{bmatrix} + \begin{bmatrix}
        1&0\\0&1
    \end{bmatrix} \otimes \begin{bmatrix}\mathbf{0} & \mathbf{0} & \mathbf{0} & \mathbf{0} \\ \mathbf{0} &\mathbf{0}&\mathbf{0}&\mathbf{0}\end{bmatrix} \\
    \mathbf{B}_1^2=\begin{bmatrix}
        \mathbf{0} & \mathbf{0} & \mathbf{0} & \mathbf{0} &\mathbf{0} & \mathbf{0} & \mathbf{0} & \mathbf{0} \\
        \mathbf{0} & \mathbf{0} & \mathbf{0} & \mathbf{0} &\mathbf{0} & \mathbf{0} & \mathbf{0} & \mathbf{0} \\
        \mathbf{0} & \mathbf{0} & \mathbf{0} & \mathbf{0} &\mathbf{0} & \mathbf{0} & \mathbf{0} & \mathbf{0} \\
        \mathbf{0} & \mathbf{0} & \mathbf{0} & \mathbf{0} &\mathbf{0} & \mathbf{0} & \mathbf{0} & \mathbf{0} \\
    \end{bmatrix},
\end{align}
and
\begin{align}
\mathbf{B}_0^3=\mathbf{B}_3^1\otimes \mathbb{I} \otimes \mathbb{I}+\mathbb{I}\otimes \mathbf{B}_3^{1}\\
\mathbf{B}_0^3= \begin{bmatrix}
    \boldsymbol{\beta}& \boldsymbol{\gamma}\\ \boldsymbol{\eta} & \boldsymbol{\kappa}
\end{bmatrix}
\otimes \begin{bmatrix}
    1&0\\0&1
\end{bmatrix}
\otimes \begin{bmatrix}
    1&0\\0&1
    \end{bmatrix}+
\begin{bmatrix}
    1&0\\0&1
    \end{bmatrix} \otimes
    \begin{bmatrix}
        2\boldsymbol{\beta}& \boldsymbol{\gamma}& \boldsymbol{\gamma}& \mathbf{0}\\
        \boldsymbol{\eta} & \boldsymbol{\beta}+\boldsymbol{\kappa}& \mathbf{0} & \boldsymbol{\gamma}\\
        \boldsymbol{\eta} & \mathbf{0} & \boldsymbol{\beta}\boldsymbol{\kappa}& \boldsymbol{\gamma}\\
        \mathbf{0} & \boldsymbol{\eta} & \boldsymbol{\eta} & 2\boldsymbol{\kappa}\\
    \end{bmatrix}\\
\mathbf{B}_0^3=\begin{bmatrix}
3\boldsymbol{\beta}& \boldsymbol{\gamma}& \boldsymbol{\gamma}& \mathbf{0} & \boldsymbol{\gamma}& \mathbf{0} & \mathbf{0} & \mathbf{0} \\
\boldsymbol{\eta} & 2\boldsymbol{\beta}+ \boldsymbol{\kappa}& \mathbf{0} & \boldsymbol{\gamma}& \mathbf{0} & \boldsymbol{\gamma}& \mathbf{0} & \mathbf{0} \\
\boldsymbol{\eta} & \mathbf{0} & 2\boldsymbol{\beta} + \boldsymbol{\kappa} & \boldsymbol{\gamma} & \mathbf{0} & \mathbf{0} & \boldsymbol{\gamma} & \mathbf{0} \\
\mathbf{0} & \boldsymbol{\eta} & \boldsymbol{\eta} & \boldsymbol{\beta} + 2\boldsymbol{\kappa} & \mathbf{0} & \mathbf{0} & \mathbf{0} & \boldsymbol{\gamma} \\
\boldsymbol{\eta} & \mathbf{0} & \mathbf{0} & \mathbf{0} & 2\boldsymbol{\beta}+ \boldsymbol{\kappa}& \boldsymbol{\gamma}& \boldsymbol{\gamma}& 0 \\
\mathbf{0} & \boldsymbol{\eta} & \mathbf{0} & \mathbf{0} & \boldsymbol{\eta} & \boldsymbol{\beta} + 2\boldsymbol{\kappa}& \mathbf{0} & \boldsymbol{\gamma}\\
\mathbf{0} & \mathbf{0} & \boldsymbol{\eta} & \mathbf{0} & \boldsymbol{\eta} & \mathbf{0} & \boldsymbol{\beta} + 2\boldsymbol{\kappa}& \boldsymbol{\gamma}\\
\mathbf{0} & \mathbf{0} & \mathbf{0} & \boldsymbol{\eta} & \mathbf{0} & \boldsymbol{\eta} & \boldsymbol{\eta} & 3\kappa
\end{bmatrix}.
\end{align}

Therefore, the Carleman matrix is 
\setcounter{MaxMatrixCols}{15}
\begin{align} \mathbf{M_3}=
\begin{bmatrix}
\boldsymbol{\beta} & \boldsymbol{\gamma}& \mathbf{0} & \mathbf{0} & \mathbf{0} & \mathbf{0} & \mathbf{0} & \boldsymbol{\xi} & \mathbf{0} & \mathbf{0} & \mathbf{0} & \mathbf{0} & \mathbf{0} & \mathbf{0} \\
\boldsymbol{\eta} & \boldsymbol{\kappa}& \mathbf{0} & \mathbf{0} & \mathbf{0} & \mathbf{0} & \mathbf{0} & \boldsymbol{\psi} & \mathbf{0} & \mathbf{0} & \mathbf{0} & \mathbf{0} & \mathbf{0} & \mathbf{0} \\
\mathbf{0} & \mathbf{0} & 2\boldsymbol{\beta} & \boldsymbol{\gamma}& \boldsymbol{\gamma}& \mathbf{0} & \mathbf{0} & \mathbf{0} & \mathbf{0} & \mathbf{0} & \mathbf{0} & \mathbf{0} & \mathbf{0} & \mathbf{0} \\
\mathbf{0} & \mathbf{0} & \boldsymbol{\eta} & \boldsymbol{\beta} + \boldsymbol{\kappa}& \mathbf{0} & \boldsymbol{\gamma}& \mathbf{0} & \mathbf{0} & \mathbf{0} & \mathbf{0} & \mathbf{0} & \mathbf{0} & \mathbf{0} & \mathbf{0} \\
\mathbf{0} & \mathbf{0} & \boldsymbol{\eta} & \mathbf{0} & \boldsymbol{\beta} + \boldsymbol{\kappa}& \boldsymbol{\gamma}& \mathbf{0} & \mathbf{0} & \mathbf{0} & \mathbf{0} & \mathbf{0} & \mathbf{0} & \mathbf{0} & \mathbf{0} \\
\mathbf{0} & \mathbf{0} & \mathbf{0} & \boldsymbol{\eta} & \boldsymbol{\eta} & 2\boldsymbol{\kappa}& \mathbf{0} & \mathbf{0} & \mathbf{0} & \mathbf{0} & \mathbf{0} & \mathbf{0} & \mathbf{0} & \mathbf{0} \\
\mathbf{0} & \mathbf{0} & \mathbf{0} & \mathbf{0} & \mathbf{0} & \mathbf{0} & 3\boldsymbol{\beta} & \boldsymbol{\gamma}& \boldsymbol{\gamma}& \mathbf{0} & \boldsymbol{\gamma}& \mathbf{0} & \mathbf{0} & \mathbf{0} \\
\mathbf{0} & \mathbf{0} & \mathbf{0} & \mathbf{0} & \mathbf{0} & \mathbf{0} & \boldsymbol{\eta} & 2\boldsymbol{\beta} + \boldsymbol{\kappa}& \mathbf{0} & \boldsymbol{\gamma}& \mathbf{0} & \boldsymbol{\gamma}& \mathbf{0} & \mathbf{0} \\
\mathbf{0} & \mathbf{0} & \mathbf{0} & \mathbf{0} & \mathbf{0} & \mathbf{0} & \boldsymbol{\eta} & \mathbf{0} & 2\boldsymbol{\beta} + \boldsymbol{\kappa}& \boldsymbol{\gamma}& \mathbf{0} & \mathbf{0} & \boldsymbol{\gamma}& \mathbf{0} \\
\mathbf{0} & \mathbf{0} & \mathbf{0} & \mathbf{0} & \mathbf{0} & \mathbf{0} & \mathbf{0} & \boldsymbol{\eta} & \boldsymbol{\eta} & \boldsymbol{\beta} + 2\boldsymbol{\kappa}& \mathbf{0} & \mathbf{0} & \mathbf{0} & \boldsymbol{\gamma}\\
\mathbf{0} & \mathbf{0} & \mathbf{0} & \mathbf{0} & \mathbf{0} & \mathbf{0} & \boldsymbol{\eta} & \mathbf{0} & \mathbf{0} & \mathbf{0} & 2\boldsymbol{\beta} + \boldsymbol{\kappa}& \boldsymbol{\gamma}& \boldsymbol{\gamma}& \mathbf{0} \\
\mathbf{0} & \mathbf{0} & \mathbf{0} & \mathbf{0} & \mathbf{0} & \mathbf{0} & \mathbf{0} & \boldsymbol{\eta} & \mathbf{0} & \mathbf{0} & \boldsymbol{\eta} & \boldsymbol{\beta} + 2\boldsymbol{\kappa}& \mathbf{0} & \boldsymbol{\gamma}\\
\mathbf{0} & \mathbf{0} & \mathbf{0} & \mathbf{0} & \mathbf{0} & \mathbf{0} & \mathbf{0} & \mathbf{0} & \boldsymbol{\eta} & \mathbf{0} & \boldsymbol{\eta} & \mathbf{0} & \boldsymbol{\beta} + 2\boldsymbol{\kappa}& \boldsymbol{\gamma}\\
\mathbf{0} & \mathbf{0} & \mathbf{0} & \mathbf{0} & \mathbf{0} & \mathbf{0} & \mathbf{0} & \mathbf{0} & \mathbf{0} & \boldsymbol{\eta} & \mathbf{0} & \boldsymbol{\eta} & \boldsymbol{\eta} & 3\boldsymbol{\kappa}\\
\end{bmatrix}
\end{align}
This represents a set of 14 linear differential equations for each spatial node. 
The eigenvalues $\lambda_i$ for $i=1, 2, 3$ are given by 
\begin{align}
    (\mathbf{B}_0^1-\boldsymbol{\lambda_1})(\mathbf{B}_0^2-\boldsymbol{\lambda_2})(\mathbf{B}_0^3-\boldsymbol{\lambda_3)}=\boldsymbol{0}.
\end{align}
As $k$ changes, the exact eigenvalues are impacted. By examining matrices $\mathbf{B}_0^1$, $\mathbf{B}_0^2$, and $\mathbf{B}_0^3$ we see that the eigenvalues depend on $\beta$, $\gamma$, $\eta$, and $\kappa$. In the GM model, $\gamma=0$ and $\eta=0$, so the eigenvalues only depend on two values, $\beta$ and $\kappa$  These terms depend on the diffusion coefficients and the linear terms. From this, we deduce that the spectral norm of $\mathbf{M}_3$ is most influenced by the diffusion coefficients and the linear terms.

\section{Runge-Kutta Methods} \label{RK4}
The Runge-Kutta methods are a family of iterative techniques for solving ordinary differential equations (ODEs). Combined with a method to discretize the spatial domain, such as finite difference method, Runge-Kutta methods can be used for PDEs. The fourth-order Runge-Kutta (RK4) method is particularly popular due to its balance between accuracy and computational efficiency \cite{shamsi2010}.

Consider the 2-species Turing initial value problem:
\begin{equation}
    \frac{d\mathbf{Y}}{dt} = D_\mathbf{Y}\nabla^2\mathbf{Y}+f(t, \mathbf{Y}), \quad \mathbf{X(t=0)} = \mathbf{X_0}.
\end{equation}
where $D_\mathbf{Y}=\begin{bmatrix}
    D_1 \\ D_2
\end{bmatrix}$. 
The RK4 method computes the solution at time step, $\mathbf{Y}_{b+1}$, from the pervious solution at time point $\mathbf{Y}_b$ using the following iterative scheme:

\begin{align}
\mathbf{k}_1 = f(t_b, \mathbf{Y}_b)
\end{align}
\begin{align}
\mathbf{k}_2 = f\left(t_b + \frac{h_t}{2}, \mathbf{Y}_b + \frac{h_t}{2} \mathbf{k}_1 \right)
\end{align}
\begin{align}
\mathbf{k}_3 = f\left(t_b + \frac{h_t}{2}, \mathbf{Y}_b + \frac{h_t}{2} \mathbf{k}_2 \right)
\end{align}
\begin{align}
\mathbf{k}_4 = f(t_b + h_t, \mathbf{Y}_b + h_t \mathbf{k}_3)
\end{align}
\begin{align}
\mathbf{Y}_{b+1} = \mathbf{Y}_b + \frac{h_t}{6} \left(\mathbf{k}_1 + 2\mathbf{k}_2 + 2\mathbf{k}_3 + \mathbf{k}_4 \right)
\end{align}
Here, $h_t$ is the time step size, and $\mathbf{k}_1, \mathbf{k}_2, \mathbf{k}_3, \mathbf{k}_4$ represent intermediate slopes used to approximate the solution, similar to the Euler method. We aim to have the error between the RK4 method and Carleman-GM approximations to be approaching computer epsilon, which is the maximum precision due to the computer processor. The computer used for the subsequent simulations has computer epsilon $10^{-10}$. We use RK4 the solve the GM model for the numerical simulations. 

\section{Spectral Norm of the Discrete Laplacian with Periodic Boundary Conditions}\label{sec:laplacian-norm}

\begin{lemma}[Spectral Norm of 1D Discrete Laplacian]
Let $\mathcal D_h \in \mathbb{R}^{n \times n}$ be the discrete Laplacian matrix with central differences and periodic boundary conditions on a uniform grid with spacing $h$ and $n$ grid points. Then the spectral norm of $L_h$ is:
$$\|\mathcal D_h\|_2 = \frac{4}{h^2}$$
which is 
\begin{equation}\label{eq:laplacian-norm-1d}
\|\mathcal D_h\|_2 = 4 n^2
\end{equation}
in terms of the gird points $n$.
\end{lemma}

\begin{proof}
The discrete Laplacian matrix $L_h$ with periodic boundary conditions is given by:
\begin{equation}\label{eqa:D_h}
\mathcal D_h = \frac{1}{h^2} \begin{bmatrix} 
-2 & 1 & 0 & \cdots & 0 & 1 \\
1 & -2 & 1 & \cdots & 0 & 0 \\
0 & 1 & -2 & \cdots & 0 & 0 \\
\vdots & \vdots & \vdots & \ddots & \vdots & \vdots \\
0 & 0 & 0 & \cdots & -2 & 1 \\
1 & 0 & 0 & \cdots & 1 & -2
\end{bmatrix}
\end{equation}

For a circulant matrix with first row $\mathbf{c} = [c_0, c_1, \ldots, c_{n-1}]$, the eigenvalues are:
$$\lambda_k = \sum_{j=0}^{n-1} c_j \omega_n^{jk}, \quad k = 0, 1, \ldots, n-1$$
where $\omega_n = e^{2\pi i/n}$ is the primitive $n$-th root of unity. 
$\mathcal D_h$ is a circulant matrix with first row $\mathbf{c} = \frac{1}{h^2}[-2, 1, 0, \ldots, 0, 1]$.
Therefore, the eigenvalues of \cref{eqa:D_h} is 
$$\lambda_k = \frac{1}{h^2}(-2 + \omega_n^k + \omega_n^{(n-1)k})$$
We now establish that $\omega_n^{(n-1)k} = \overline{\omega_n^k}$, where $\overline{z}$ denotes the complex conjugate.
Since $\omega_n^{(n-1)k} = e^{2\pi i (n-1)k/n} = e^{2\pi i k - 2\pi i k/n} = e^{2\pi i k} \cdot e^{-2\pi i k/n}$ and $e^{2\pi i k} = 1$ for any integer $k$:
$$\omega_n^{(n-1)k} = e^{-2\pi i k/n} = \overline{e^{2\pi i k/n}} = \overline{\omega_n^k}$$
Here, $\omega_n^k = e^{2\pi i k/n} = \cos(2\pi k/n) + i\sin(2\pi k/n)$ is a point on the unit circle at angle $2\pi k/n$,
$\overline{\omega_n^k} = \cos(2\pi k/n) - i\sin(2\pi k/n)$ is its reflection across the real axis.
For any complex number $z$, we have $z + \overline{z} = 2\text{Re}(z)$.
Therefore:
\begin{align}
\lambda_k &= \frac{1}{h^2}(-2 + \omega_n^k + \overline{\omega_n^k}) \\
&= \frac{1}{h^2}(-2 + 2\text{Re}(\omega_n^k)) \\
&= \frac{1}{h^2}(-2 + 2\cos(2\pi k/n)) \\
&= \frac{2}{h^2}(\cos(2\pi k/n) - 1)
\end{align}

Since $-1 \leq \cos(2\pi k/n) \leq 1$ for all $k$:
\begin{enumerate}
\item When $k = 0$: $\cos(0) = 1 \Rightarrow \lambda_0 = 0$
\item When $\cos(2\pi k/n) = -1$: $\lambda_k = -\frac{4}{h^2}$
\end{enumerate}

The condition $\cos(2\pi k/n) = -1$ occurs when $2\pi k/n = \pi + 2\pi m$ for integer $m$, i.e., when $k = n/2$ (if $n$ is even) or $k = (n \pm 1)/2$ (if $n$ is odd).

The spectral norm is:
$$\|\mathcal D_h\|_2 = \max_{k=0,\ldots,n-1} |\lambda_k| = \max_{k=0,\ldots,n-1} \left|\frac{2}{h^2}(\cos(2\pi k/n) - 1)\right|$$

Since $\cos(2\pi k/n) - 1 \leq 0$ for all $k$:
$$|\lambda_k| = \frac{2}{h^2}(1 - \cos(2\pi k/n))$$

The maximum occurs when $\cos(2\pi k/n) = -1$:
$$\|\mathcal D_h\|_2 = \frac{2}{h^2}(1 - (-1)) = \frac{4}{h^2}$$
\end{proof}

\begin{corollary}[Spectral Norm of Multi-dimensional Discrete Laplacian]
Let $\mathcal D_h^{(d)} \in \mathbb{R}^{N \times N}$ be the $d$-dimensional discrete Laplacian with central differences and periodic boundary conditions on a uniform grid with spacing $h$ and $n_i$ grid points in direction $i$, where $N = \prod_{i=1}^d n_i$. Then:
$$\|\mathcal D_h^{(d)}\|_2 = \frac{4d}{h^2}$$
which is 
\begin{equation}\label{eq:laplacian-norm}
  \|\mathcal D_h^{(d)}\|_2 = 4d n^2
\end{equation}
in terms of the gird points $n$ in 1D.
\end{corollary}

\begin{proof}

The $d$-dimensional discrete Laplacian can be written as:
$$\mathcal D_h^{(d)} = \sum_{i=1}^d I_1 \otimes \cdots \otimes I_{i-1} \otimes L_h^{(1)} \otimes I_{i+1} \otimes \cdots \otimes I_d$$
where $\mathcal D_h^{(1)}$ is the 1D Laplacian and $I_j$ are identity matrices of appropriate dimensions.

The eigenvalues of $\mathcal D_h^{(d)}$ are:
\begin{align}
\lambda_{k_1,\ldots,k_d} &= \sum_{i=1}^d \lambda_{k_i}^{(1)} = \sum_{i=1}^d \frac{2}{h^2}(\cos(2\pi k_i/n_i) - 1) \\
&= \frac{2}{h^2}\sum_{i=1}^d (\cos(2\pi k_i/n_i) - 1)
\end{align}

$$|\lambda_{k_1,\ldots,k_d}| = \frac{2}{h^2}\left|\sum_{i=1}^d (\cos(2\pi k_i/n_i) - 1)\right|$$

Since $\cos(2\pi k_i/n_i) - 1 \leq 0$ for all $i$:
$$|\lambda_{k_1,\ldots,k_d}| = \frac{2}{h^2}\sum_{i=1}^d (1 - \cos(2\pi k_i/n_i))$$

The maximum occurs when $\cos(2\pi k_i/n_i) = -1$ for all $i$:
$$\|\mathcal D_h^{(d)}\|_2 = \frac{2}{h^2}\sum_{i=1}^d (1 - (-1)) = \frac{2}{h^2} \cdot 2d = \frac{4d}{h^2}$$
\end{proof}

\end{document}